\documentclass[onecolumn,12pt,journal]{IEEEtran}

\usepackage{xcolor}
\usepackage{cite}
\usepackage[OT1]{fontenc}
\usepackage{amsmath}
\usepackage{subeqnarray} 
\usepackage{amssymb}
\usepackage{graphicx}
\usepackage{array}
\usepackage{multicol, multirow}
\usepackage{psfrag}

\usepackage{subfigure}
\usepackage{amssymb}
\usepackage{color}   
\usepackage{epsf}
\usepackage{algorithmic}
\usepackage{algorithm}
\usepackage{url}
\usepackage{setspace}
\usepackage[margin=2.5cm]{geometry}

\begin{document}

\title{An Analytical Model for the Intercell Interference Power in the Downlink of Wireless Cellular Networks}

\author{Benoit Pijcke, Marie Zwingelstein-Colin, Marc Gazalet, \\ Mohamed Gharbi, Patrick Corlay \\
Universit\'{e} Lille Nord de France, F-59000 Lille \\
UVHC, IEMN/DOAE, F-59313 Valenciennes\\
CNRS, UMR 8520, F-59650 Villeneuve d'Ascq, France\\
\selectfont\ttfamily
\itshape firstname.lastname@univ-valenciennes.fr\\ }


\maketitle

\begin{abstract}
In this paper, we propose a methodology for estimating the statistics of the intercell interference power in the downlink of a multicellular network. We first establish an analytical expression for the probability law of the interference power when only Rayleigh multipath fading is considered. Next, focusing on a propagation environment where small-scale Rayleigh fading as well as large-scale effects, including attenuation with distance and lognormal shadowing, are taken into consideration, we elaborate a semi-analytical method to build up the histogram of the interference power distribution. From the results obtained for this combined small- and large-scale fading context, we then develop a statistical model for the interference power distribution. The interest of this model lies in the fact that it can be applied to a large range of values of the shadowing parameter. The proposed methods can also be easily extended to other types of networks. 
\end{abstract}


\section{Introduction}

In the emerging wireless communication standards LTE-Advanced and Mobile WiMAX, aggressive spectrum reuse is mandatory in order to achieve the increased spectral efficiency required by IMT-Advanced for the 4th generation of standard telephony. However, since spectrum reuse comes at the expense of increased intercell interference, these standards explicitely require interference management as a basic system functionality \cite{Him_CM10,80216mSDD,3GPP_36300}. The research area related to the development and analysis of interference management techniques, mostly in relation with the more general subject of radio ressource management, is very dynamic, as witnessed by the high number of relevant recent contributions in this area \cite{Choi_WCN10, Bou_CM09, Zhang_TJCUPT10, Hernandez_WCMC10, Kang_WCNC09, Gesbert_JSAC10, Kiani_TWC08}. All these new standards use OFDMA as the modulation and the multiple access scheme. In an OFDMA system, there is no intracell interference as the users remain orthogonal, even through multipath channels. However, when users from different cells are present at the same time on the same subchannel, which is the case under aggressive frequency reuse, signals superpose, leading to some form of intercell interference.

Providing statistical models of the interference power is essential to allow for an accurate evaluation of networks performances without the need for lenghtly and costly Monte Carlo simulations. The statistical characterization of the interferences has been investigated for a long time, under lots of different scenarios, and following several approaches. The distribution of cumulated instantaneous interference power in a Rayleigh fading channel was investigated in \cite{Mathar_WN95}, where an infinite number of interfering stations was considered. In \cite{Zorzi_TC97}, the interference power statistics is obtained analytically for the uplink and downlink of a cellular system, but in the presence of large-scale fading only. Interference modeling when considering only large-scale fading effects has also been investigated in \cite{Beau_TVT04, San_EL06,Szy_PhD07}, where the emphasis is on finding a good approximation of the lognormal sum distribution. In \cite{Haas_CL04}, an analytical derivation of the probability density function (pdf) of the adjacent channel interference is derived for the uplink. More recently, in \cite{Sung_JWCN10} the pdf of the downlink SINR was derived in the context of randomly located femtocells via a semi-analytical method. Other contributions have focused directly on the analysis of a particular performance measure that is influenced by intercell interference, like the probability of outage and the radio spectrum efficiency \cite{Pra_TVT91, Berggren_CL04, Pratesi_ISPIMRC06}. The analysis of interference in dense asynchronous networks, such as ad-hoc networks, is also an active research area, for which a deep review of the recent developments can be found in \cite{Hae_Now08, Car_CST10}.

In this paper, we derive a semi-analytical methodology to estimate the statistics of the intercell interference power in a wireless cellular network, when the combined effects of large-scale and small-scale multipath fading are taken into consideration. Large-scale effects include attenuation with distance (path-loss) as well as lognormal shadowing, and the small-scale fading is Rayleigh distributed. We consider a distributed wireless multicellular network, in both cases where power control and no power control is applied. The proposed methodology is semi-analytical, in that the statistical estimate of the interference power resulting from $N>1$ interferers is obtained by numerical techniques from an analytically-derived interference model for one interferer. The methodology is valid in a quite general framework; we have chosen to present it using a hexagonal network layout, although it can handle any other topology. We validate the proposed methods by comparing the moments of the estimates to the exact moments of the distribution which can be derived analytically. Using this methodology, we are able to provide a very good estimate of the pdf of the interference power, for different values of the shadowing standard deviation, $\sigma_\mathrm{dB}$. Based on these estimates, we then propose an analytical statistical model of the interference power, based on a modified Burr distribution, which includes 5 parameters. This analytical, parameterized by~$\sigma_\mathrm{dB}$, model will hopefully serve as a practical tool for the assesment and simulation of wireless cellular networks when the effect of shadowing is to be considered.

The main contributions of this paper are as follows.
\begin{itemize}
\item In the special situation where only path-loss and Rayleigh fading are considered (no shadowing), we derive a very accurate approximated analytical expression for the pdf and the cumulative distribution function (cdf) of the intercell interference power; 
\item We propose a semi-analytical method for the estimation of the pdf of the intercell interference power in a multicellular network when the combined propagation effects of path-loss, Rayleigh fading and lognormal shadowing are considered; 
\item Based on this method, we derive an analytical model for the pdf of the intercell interference power by slightly modifying a Burr probability distribution. This model is parameterized by the lognormal standard deviation $\sigma_{\text{dB}}$ and its interest resides in the fact that it is valid on the whole $[0,12]$-dB range of values. 
\end{itemize}

The remainder of this paper is organized as follows. In Section \ref{System_model}, we describe the multicell downlink transmission environment, and we provide the expression of the interference power for which we want to find a statistical model. In Section \ref{methodology}, the original methodology for estimating the statistics of the interference power is presented. For this purpose, we examine in Section \ref{fading_only} the particular case where path loss and Rayleigh fast-fading are the only fading phenomena considered. In Section \ref{fading_shadowing}, we include the shadowing effect and we consider in the first instance the contribution of one interfering cell. We then generalize to $N>1$ interferers. In Section \ref{results}, we apply the proposed method to estimate the pdf of the interference power in a typical multicellular network, under two frequency reuse scenarios. Section \ref{model} is dedicated to the parametric analytical modeling of the interference power. Section \ref{Conclusion} concludes the paper by summarizing the proposed methods and by presenting some perspectives.

We will use the following notation for the rest of the paper. Non-bold letters such as $x$ are used to denote scalar variables, and $|x|$ is the magnitude of $x$. Bold letters like $\mathbf{x}$ denote vectors. We use $\mathbb{E}\left\{X\right\}$ to denote the expectation of~$X$. The pdf and cdf of the random variable (r.v.)~$X$ will be denoted $p_X\left(x\right)$ and $F_X\left(x\right)$ respectively.

\section{\label{System_model}Multicell downlink transmission model}

\begin{figure}
\noindent \begin{centering}
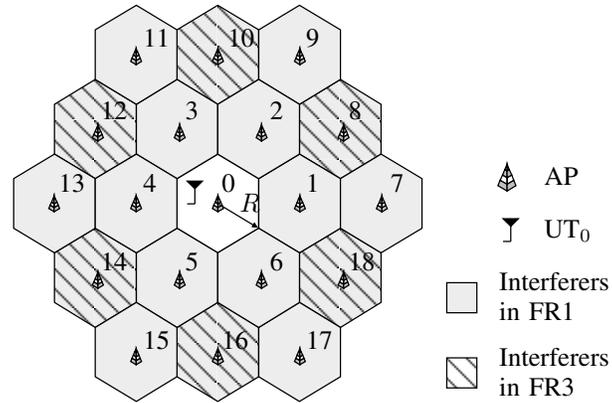
\par\end{centering}
\caption{\label{fig_network_model}Hexagonal model for a 19-cell cellular network. The largest distance from a user to its serving AP is denoted~$R$. We study the interference power undergone by the mobile receiver $\textnormal{UT}_0$ in the central cell (numbered~$0$).}
\end{figure}

We consider the downlink of an OFDMA-based 19-cell cellular network having the 2D hexagonal layout depicted on Fig.~\ref{fig_network_model}. We assume a unit-gain omnidirectional SISO (\emph{single input, single output}) antenna pattern, both for the fixed access points (APs) and the mobile user terminals (UTs) which are supposed to be uniformly distributed over the service area. As OFDM is used for intracell communication, we assume an orthogonal transmission scheme within a cell. We consider a synchronous discrete-time communication model in which active APs at any given time slot send information symbols to their respective UTs over a shared spectral resource, which gives rise to an interference-limited environment. In this framework, we will focus on the statistics of the so-called intercell interference power undergone by a typical UT. In this regard, we will consider UT in cell~$0$ (denoted $\textnormal{UT}_0$, see Fig.~\ref{fig_network_model}), for it is surrounded by 18 potential interferers. For $\textnormal{UT}_0$, the received signal on OFDMA subchannel $\ell$ at time slot $m$ can be modeled as 
\[
y_0(m,\ell)=h_0(m,\ell)x_0(m,\ell)+\sum_{n=1}^{N}h_n(m,\ell)x_n(m,\ell)+w(m,\ell).
\]
Here $x_0\left(m,\ell\right)$ represents the information symbol intended to $\textnormal{UT}_0$ and $x_n\left(m,\ell\right)$, $n \neq 0$, the $n$th interfering symbol (this symbol is sent from AP $n$ to its respective user). The coefficient $h_n\left(m,\ell\right)$ denotes the instantaneous gain of the $\ell$th (interfering) subchannel from AP~$n$ to $\textnormal{UT}_0$. Each subchannel $\ell$ is subject to additive white Gaussian noise $w\left(m,\ell\right)$. In the following, we will focus without loss of generality on a single OFDMA subchannel, thereby omitting subchannel index $\ell$ in all subsequent notations. 

Two frequency reuse scenarios will be considered (see Fig.~\ref{fig_network_model}): 
\begin{itemize}
\item the full frequency reuse pattern, denoted FR1, where all APs in the network transmit at the same time using the same frequency range ($N=18$ intercell interferers); 
\item a partial frequency reuse pattern, denoted FR3, with reuse factor~3 ($N=6$ interferers). 
\end{itemize}

Each channel is assumed to be flat-fading, possibly experiencing small-scale multipath fading and/or large-scale effects. For the rest of the paper, we concentrate on the instantaneous \emph{channel power gain}\footnote{As this paper will focus on \emph{power} gains only, the term \emph{power} will then be omitted in subsequent paragraphs.} $G_n\left(r_n\right)$ , which is proportional to~$\left|h_n\left(m\right)\right|^2$ and can be expressed as a three-factor product:
\begin{equation}
G_n\left(r_n\right)=G_{\textnormal{pl},n}\left(r_n\right) G_{\textnormal{f},n} G_{\textnormal{s},n}, n=1,2,\dots,N. 
\label{Gnrn_Gplnrn_}
\end{equation}
In the above equation, $r_n$ denotes the distance between $\textnormal{UT}_0$ and AP~$n$ (distances $r_n$ are functions of $\textnormal{UT}_0$'s position within its cell). $G_{\textnormal{pl},n}\left(r_n\right)=K\left(1/r_n\right)^\gamma$ is the (deterministic) path loss (normalized with distance, see Appendix~\ref{appendix0}), where $K$~is a constant and $\gamma$ represents the path loss exponent. The Rayleigh fading gain $G_{\textnormal{f},n}$ is modeled by an exponential distribution with rate parameter equal to~$1$, i.e., $\mathbb{E}\left\{G_{\textnormal{f},n}\right\}=1$; we denote the corresponding pdf by $p_{G_{\textnormal{f},n}}\left(x\right)$. The shadowing gain $G_{\textnormal{s},n}$ is modeled by a lognormal distribution whose pdf can be written
\[
p_{G_{\textnormal{s},n}}\left(x\right)=\frac{\xi}{\sqrt{2\pi}\sigma_{\textnormal{dB}}x}\exp \left(-\dfrac{\left(10\log_{10}\left(x\right)-\mu_{\textnormal{dB}}\right)^{2}}{2\sigma_{\textnormal{dB}}^{2}}\right),x>0,
\]
where $\xi=10/\ln(10)$ \cite{Goldsmith2005}. Note that the importance of the shadowing phenomenon is directly related to the standard deviation $\sigma_{\textnormal{dB}}$. For a given $\sigma_{\textnormal{dB}}$, the parameter $\mu_{\textnormal{dB}}$ is determined to ensure a unit mean shadowing gain: $\mathbb{E}\left\{G_{\textnormal{s},n}\right\}=1$, which leads to $\mu_{\textnormal{dB}}=-\sigma_{\textnormal{dB}}^2/\left(2\xi\right)$. As r.v.'s $G_{\textnormal{f},n}$ and $G_{\textnormal{s},n}$ are independent from each other, and as $\mathbb{E}\left\{G_{\textnormal{f},n}\right\} = \mathbb{E}\left\{G_{\textnormal{s},n}\right\} = 1$, we have, from~\eqref{Gnrn_Gplnrn_}, $\mathbb{E}\left\{G_n\left(r_n\right)\right\}=G_{\textnormal{pl},n}\left(r_n\right)$, which reflects the fact that the $n$th interfering channel's Rayleigh fading and shadowing components cause the actual gain $G_n\left(r_n\right)$ to fluctuate about its mean value $G_{\textnormal{pl},n}\left(r_n\right)$. 

The total interference power undergone by $\textnormal{UT}_0$ can then be written $I=\sum_{n=1}^{N}P_n G_n(r_n)$, where $P_n=\mathbb{E}\left\{ \left|x_n\right|^2 \right\}$ is the power emitted by AP~$n$. 
In what follows, we consider that all APs transmit at the same power, i.e., $P_n=P$ for all $n$. This corresponds to, e.g., a fast-fading environment where no channel state information feeds back from mobile users to APs, which results in a no power control scheme where all APs transmit at the maximum power; although crude, this scheme can be seen as a lower bound on performance for real systems. Considering that each AP transmits at the same power~$P$ also applies to a more practical scenario where APs have access to channel state information and power control is associated with the opportunistic scheduling policy proposed (and proved to be sum-rate optimal) in \cite{Kiani_TWC08}, when the number of users per cell is high (since in this case, it can be expected that the channels between users scheduled at the same time and their serving APs have about the same power gains). Thus, the interference simplifies to $I=P\sum_{n=1}^{N}G_{n}\left(r_{n}\right)$. 

We now define the \emph{interference gain} ---~which will be denoted~$G$~--- as being the sum of the channel power gains between the interested user and the $N$~interferers, i.e., 
\begin{equation}
G =\sum_{n=1}^{N}G_n(r_n) =\sum_{n=1}^{N}G_{\textnormal{pl},n}(r_n)G_{\textnormal{f},n}G_{\textnormal{s},n}.
\label{G_sumn1N_Gplnrn_Gfn_Gsn}
\end{equation}
(Note that $G$ is a function of $\textnormal{UT}_{0}$'s location through the distances~$r_n$.) So, as $I=P G$, characterizing the interference power~$I$ is equivalent to studying the interference gain~$G$. We will concentrate on the latter in the subsequent sections.

\section{\label{methodology}Methodology}
We are now interested in finding an estimate of the pdf of the random interference gain~\eqref{G_sumn1N_Gplnrn_Gfn_Gsn}. Since direct calculation of the pdf does not seem possible, we aim at producing an accurate histogram for the interference gain $G$ that will then be modeled using a specified statistical distribution. Such a histogram is constructed from a set of samples called a \emph{typical} set, i.e., a discrete ensemble of values that accurately represents a random phenomenon. Traditionally (and especially in the telecommunications area), this typical set is issued from Monte Carlo simulations, which might, at first sight, produce satisfying results. However, in a propagation environment that is subject to intense shadowing (i.e., for large values of the $[0,12]$-dB range under consideration), the classical Monte Carlo method fails at producing a representative set of sampled gains \cite{Huang2004,Asmussen2000}. This can be explained by examining the particular distribution involved, for one single as well as for multiple interfering cells. A typical cdf of the interference gain (single or multiple interferers) for a high value of~$\sigma_{\textnormal{dB}}$ belongs to the class of \emph{heavy-tailed} distributions~\cite{Markovitch2007}, for which the least-frequently occuring values ---~also called \emph{rare events}~--- are the most important ones, as a proportion of the total population, in terms of moments. A finite-time random drawing process performed on this cdf never produces these rare events because of their very low probabilities, which causes the resulting set to be not typical. Hence the need for a new approach. 

As will be seen in Subsection \ref{one_interferer}, the pdf and the cdf of the interference gain for one single interferer may be expressed in its integral form. From this expression, we propose the following two-step approach: 
\begin{enumerate}
\item Produce a typical set of gains for one interferer using the \emph{generalized inverse method}. This method consists in generating a typical set of samples corresponding to an arbitrary continuous cdf~$F$, and is based upon the following property: if~$U$ is a uniform~$[0,1]$ r.v., then $F^{-1}\left(U\right)$ has cdf~$F$; 
\item Produce a typical set for multiple interferers by adequately combining typical sets from single interferers and the Monte Carlo computational technique. 
\end{enumerate}

\subsection{\label{fading_only}Special case: No shadowing}

\begin{figure}[t]
\begin{center}
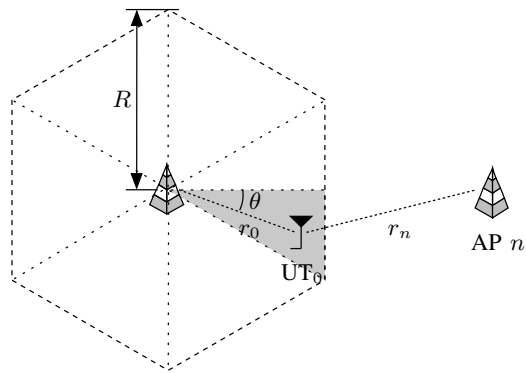
\end{center}
\caption{\label{network_symmetry}Because of the particular symmetry of the network geometry, we need only study the interference gain distribution for a user located within one the twelve dashed triangular areas. For illustration purposes, we will consider the grey-shaded sector.}
\end{figure}

We start this section by considering a propagation environment in which the only fading phenomenon is due to Rayleigh multipath fading. In this particular case, \eqref{Gnrn_Gplnrn_} simplifies to 
\begin{equation}
G_n\left(r_n\right)=G_{\textnormal{pl},n}\left(r_n\right) G_{\textnormal{f},n}.
\label{Gnrn_Gplnrn_Gfn}
\end{equation}
We first note that, because of the symmetry of the network geometry, we need only study the interference power distribution for $\textnormal{UT}_0$ located within one of the twelve triangular sectors depicted on Fig.~\ref{network_symmetry}; in the following, we will consider the grey-shaded region for illustration purposes. 

We now introduce an original approximation that will help simplify further computations. We can see that in~\eqref{Gnrn_Gplnrn_Gfn}, it is $\textnormal{UT}_0$'s random position that makes the path loss $G_{\textnormal{pl},n}\left(r_{n}\right)$ fluctuate, when the randomness of $G_{\textnormal{f},n}$ is due to Rayleigh fading. But it is worth noting that, although both phenomena are random, path loss fluctuations differ from multipath fading in an important way : the path loss takes values in a finite set (related to $\textnormal{UT}_0$'s location within its cell) whereas the variations due to fading have an (theoretically) infinite dynamic range. Since pathloss fluctuations' dynamics are very small compared to fading's, we propose to approximate~\eqref{Gnrn_Gplnrn_Gfn} by replacing each gain $G_{\textnormal{pl},n}\left(r_{n}\right)$ by its average value, which leads to 
\begin{align}
G_{n} & \approx\underset{r_{n}}{\mathbb{E}}\left\{ G_{\textnormal{pl},n}\left(r_{n}\right)\right\} G_{\textnormal{f},n}\nonumber \\
 & =\underset{r_{0},\theta}{\mathbb{E}}\left\{G_{\textnormal{pl},n}\left(f_{n}\left(r_{0},\theta\right)\right)\right\} G_{\textnormal{f},n},
\label{Gn_E_r0_theta_Gpln}
\end{align}
using the notation $r_n=f_n\left(r_0,\theta\right)$, $n=1,2,...,N$, where $\left(r_0,\theta\right)$ are $\textnormal{UT}_0$'s polar coordinates, as depicted in Fig.~\ref{network_symmetry}. By examining~\eqref{Gn_E_r0_theta_Gpln}, we see that, under this approximation, $G_n$ does not depend on $\textnormal{UT}_0$'s varying position anymore. 

We further note that $G_{n}$, as expressed in~\eqref{Gn_E_r0_theta_Gpln},
is an exponentially distributed r.v. with rate parameter~$1/\lambda_n$ \cite{Johnson1994}, $\lambda_n$ ---~which we call the \emph{average path loss}~--- being defined as follows: 
\begin{equation}
\lambda_{n}=\underset{r_{0},\theta}{\mathbb{E}}\left\{ G_{\textnormal{pl},n}\left(f_{n}\left(r_{0},\theta\right)\right)\right\}.
\label{lambdan_Gpln_Er}
\end{equation}
Using~\eqref{lambdan_Gpln_Er}, \eqref{Gn_E_r0_theta_Gpln} can also be written
\begin{equation}
G_n\approx\lambda_nG_{\textnormal{f},n},
\label{Gn_lambdan_Gfn}
\end{equation}
and the intercell interference gain~\eqref{G_sumn1N_Gplnrn_Gfn_Gsn} can be reduced to a sum of independant (but not identically distributed) exponential r.v.'s: 
\begin{equation}
G\approx\sum_{n=1}^{N}\lambda_{n}G_{\textnormal{f},n}.
\label{G_sum_lambdan_Gfn}
\end{equation}
$G$, as expressed in~\eqref{G_sum_lambdan_Gfn}, is a r.v. whose cdf, denoted $F_G\left(g\right)$, has a closed form expression available in the literature~\cite{Amari1997}; it can be expressed as 
\begin{equation}
F_G\left(x\right)=1-\sum_{n=1}^{N}A_{n}\exp\left(-\dfrac{x}{\lambda_{n}}\right),
\label{FG_1-sum_An_exp_-g_lambdan}
\end{equation}
where
\[
A_n=\dfrac{\lambda_{n}^{N}}{\overset{N}{\underset{j\neq n}{\underset{j=1}{\prod}}}\lambda_{n}-\lambda_{j}},\,\, n=1...N.
\]
The pdf, denoted~$p_G\left(g\right)$, can be easily calculated by deriving~\eqref{FG_1-sum_An_exp_-g_lambdan}:
\begin{equation}
p_G\left(x\right)=\sum_{n=1}^{N}\dfrac{A_n}{\lambda_{n}}\exp\left(-\dfrac{x}{\lambda_{n}}\right).\label{pG_sum_An_over_lambdan_}\end{equation}

In Section~\ref{results_fading}, it is first shown that approximation~\eqref{Gn_E_r0_theta_Gpln} is valid in the case of one single interfering cell. This consequently validates the proposed model~\eqref{G_sum_lambdan_Gfn} in the case of multiple interfering cells, which we show for both frequency reuse patterns FR1 and FR3. 

\subsection{\label{fading_shadowing}General case: Attenuation with distance, shadowing and multipath fading}

Let us now focus on characterizing the distribution of the intercell interference gain~$G$ in a propagation environment where Rayleigh fading as well as shadowing (due to obstacles between the transmitter and receiver that attenuate signal power) are taken into account. To the best of our knowledge, no closed form expression for the interference gain $G$ exists in the literature. But, as will be seen in Section~\ref{one_interferer}, we determine an analytical formula (under integral form) of the distribution of the interference gain for one interferer. Using this result, we are able to obtain a histogram for G's distribution in the presence of multiple interferers. 

For this purpose, we proceed in two steps: first, we compute a typical set for the interference gain produced by one single interferer. As described in Section~\ref{one_interferer}, this is done by numerical computation (from the integral-form cdf), followed by non-uniform partitioning, and then inversion, of the cdf. Then we generate a typical set for $N$ interferers using an appropriate combination of the (weighted by $\lambda_n$) typical sets of each single interferer (Section~\ref{N_interferers}). The accuracy of the proposed method will be evaluated in both single- and multiple-interferer cases by comparing the actual moments computed from the typical sets with the exact moments of the interference gain distribution (which can be formulated analytically, as will be seen in Section~\ref{preliminaries}).

\subsubsection{\label{preliminaries}Preliminaries}
We begin this section by examining two important points. 

When taking into account multipath fading as well as shadowing as the fading effects in the propagation environment, a question arises about the validity of the original approximation~\eqref{Gn_lambdan_Gfn}. Fortunately, our approximation is being strengthened by this additional contribution due to shadowing, since this phenomenon is just another source of infinite-dynamics randomness. Taking shadowing into consideration amounts to introducing an additional term in~\eqref{Gn_lambdan_Gfn} that can now be written
\begin{equation}
G_n\approx\lambda_nG_{\textnormal{f},n}G_{\textnormal{s},n}.
\label{Gn_lambdan_Gfn_Gsn}
\end{equation}

A second point pertains to the moments of both statistical distributions of~$G_n$ (single interferer) and~$G$ (multiple interferers). Using approximation~\eqref{Gn_lambdan_Gfn_Gsn}, it is shown in~Appendix~\ref{appendix1} that the $k$th-order moment of $G_n$'s distribution has the following expression:
\begin{equation}
\mathbb{E}\left\{\left(G_n\right)^k\right\}=k!\exp\left(k\left(k-1\right)\frac{\sigma_{\textnormal{dB}}^2}{2} \right).
\label{muk_k_exp_k_}
\end{equation}
Computation of the $k$th-order moment of $G$'s distribution is done in~Appendix~\ref{appendix2} and leads to the following formula:
\begin{equation}
\mathbb{E}\left\{G^k\right\} =k!\,\sum_{\mathbf{a}:|\mathbf{a}|=k}\lambda^{\mathbf{a}}\exp\left(\dfrac{\sigma_{\textnormal{dB}}^{2}}{2}\left(-k+\sum_{n=1}^{N} \alpha_{n}^{2}\right)\right),
\label{EZk_factk_sum_}
\end{equation}
where $\mathbf{a}=\left(\alpha_1,\alpha_2,\dots,\alpha_N\right)$, $\alpha_n\in \mathbb{N}$, $n=1,2,\dots,N$, is an $N$-dimensional vector whose sum of components is written $|\mathbf{a}|=\sum_{n=1}^{N}\alpha_n$, and $\lambda^{\alpha}=\lambda_{1}^{\alpha_1} \lambda_{2}^{\alpha_2} \dots \lambda_{N}^{\alpha_N}$.
So the summation in Eq~\eqref{EZk_factk_sum_} is taken over all sequences of non-negative integer indices~$\alpha_1$ through~$\alpha_N$ such that the sum of all~$\alpha_n$ is $k$. Note that the $1$st-order moment,
\begin{equation}
\mathbb{E}\left\{G\right\} = \sum_{n=1}^{N}\lambda_n,
\label{nu1}
\end{equation}
is a quantity of particular interest because it is proportional to the \emph{average} power of the interference signal. 

As closed form expressions of moments have been determined, they may be used in evaluating the accuracy of typical sets for both single- and multiple-interferer statistical laws. 

\subsubsection{\label{one_interferer}Single interferer}

We now turn on to computing a typical set for the interference gain produced by one interferer. For convenience, the average path loss~\eqref{lambdan_Gpln_Er} for this single interferer is normalized to~1, i.e., $\lambda_n=1$, so~\eqref{Gn_lambdan_Gfn_Gsn} reduces to
\begin{equation}
G_n\approx G_{\textnormal{f},n}G_{\textnormal{s},n}.
\label{Gn_Gfn_Gsn}
\end{equation}
As $G_n$ is the product of two independent r.v.'s, its cdf can be written 
\begin{align}
F_{G_n}\left(x\right) & =\int_{0}^{\infty} p_{G_{\textnormal{f},n}}\left(u\right)\left[\int_{0}^{\frac{x}{u}} p_{G_{\textnormal{s},n}} \left(y\right)\textnormal{d}y\right]\textnormal{d}u 
\nonumber \\
& =\int_{0}^{\infty}p_{G_{\textnormal{f},n}}\left(u\right) F_{G_{\textnormal{s},n}}\left(\frac{x}{u}\right) \textnormal{d}u, \label{FGnx_int_}
\end{align}
where $F_{G_{\textnormal{s},n}}\left(x/u\right)$ denotes the shadowing gain's cdf. Recalling that $G_{\textnormal{s},n}$ is modeled as a lognormal r.v., we have, using the same notations as in Section~\ref{System_model},
\[ F_{G_{\textnormal{s},n}}\left(x/u\right)=Q\left(\left(\mu_{\textnormal{dB}}-10\log\left(\frac{x}{u}\right) \right) /\sigma_{\textnormal{dB}} \right),
\]
where $Q\left(z\right)=1/\sqrt{2\pi}\int_{z}^{\infty}\exp\left(-t^{2}/2\right)\mathrm{d}t$ is the complementary error function of Gaussian statistics. Replacing $p_{G_{\textnormal{f},n}}\left(u\right)$ and $F_{G_{\textnormal{s},n}}\left(x/u\right)$
by their respective expression in~\eqref{FGnx_int_}, we obtain an integral-form expression for the cdf of the intercell interference gain produced by one single interfer: 
\begin{equation}
F_{G_n}\left(x\right)=\int_{0}^{\infty}Q\left(\dfrac{10\log_{10}\left(\frac{u}{x}\right)}{\sigma_{\textnormal{dB}}}-\dfrac{\sigma_{\textnormal{dB}}}{2\xi}\right)\exp\left(-u\right)\textnormal{d}u.
\label{FGnx_int0Infty_}
\end{equation}

We are now interested in generating a typical set of the interference gain $G_n$; we denote this typical set by $\mathcal{S}_n^\ell$, where $\ell$ is the number of elements in the set. It was mentioned in Section~\ref{methodology} that, though widely used in telecommunications, the Monte Carlo computational technique proves inefficient for large values of $\sigma_{\textnormal{dB}}$. An interesting alternative method is the generalized inverse method, for which an $\ell$-element typical set for a given distribution is obtained by an $\ell$-level uniform partitioning, followed by inversion, of the cdf. Now we know that, for large values of~$\sigma_{\textnormal{dB}}$, the distribution of $G_n$ exhibits the heavy-tailed property, which means, as described before, that the least-frequently occuring values (i.e., the highest gains) are the most important ones in terms of moments. Therefore, taking these highest amplitudes into consideration using the 'classical' generalized inverse method would require a finer partitioning of the cdf, which would produce a typical set made up of a huge amount of elements. 

\begin{figure}[t]
\begin{center}
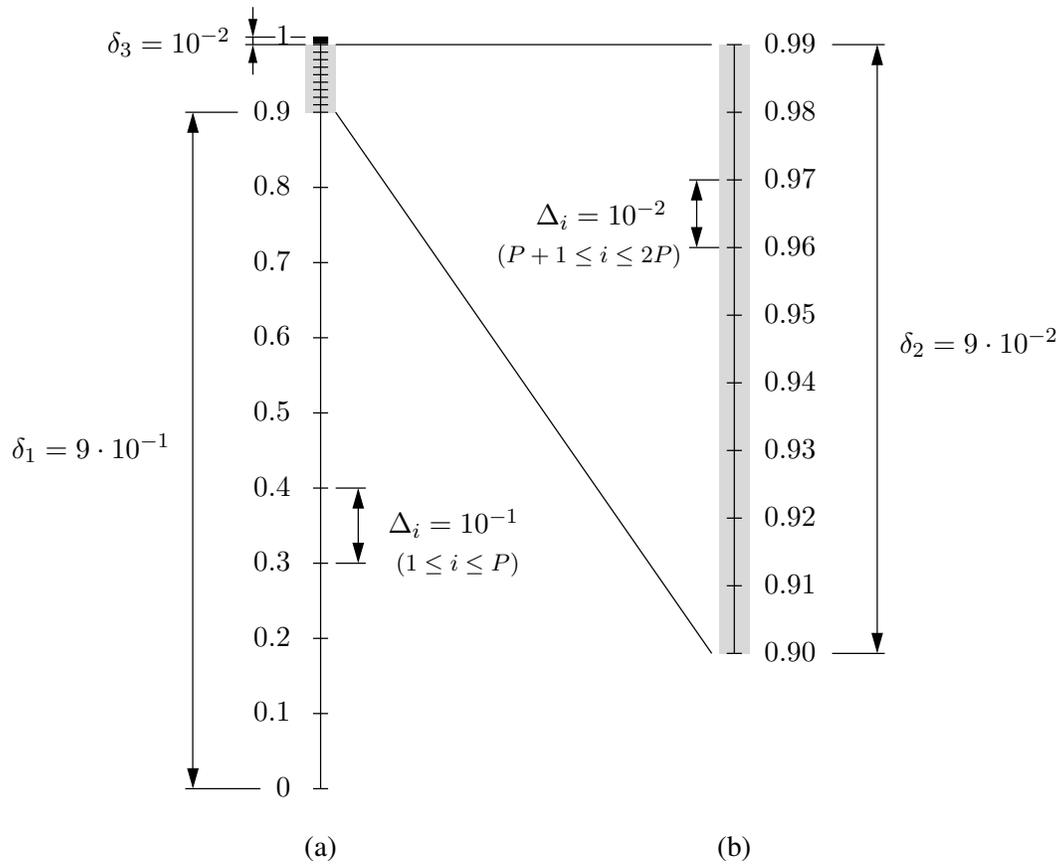
\end{center}
\caption{\label{principle_MCP}Illustration of the general inverse method with non-uniform partitioning ($J=3$, $P=9$): (a) non-uniform partitioning of the $[0,1]$ segment; (b) uniform partitioning of interval~$I_2$.}
\end{figure}

In order to construct a typical set with a reasonable value for $\ell$, we propose to accomodate the above-mentioned method by performing a \emph{non-uniform} partitioning of $G_n$'s cdf, and, as high amplitudes are important in terms of moments, we proceed with a finer partitioning of the $\left[0,1\right]$ segment for values close to~$1$. The implementation details of the method are described on Fig.~\ref{principle_MCP}; they result from a good compromise between accuracy and simplicity. We first divide the interval $\left[0\,\,1\right]$ of the cdf into $J$ intervals, numbered $j=1,\dots,J$, of different lengths: the $j$th interval has a length $d_j=9\times10^{-j}$, $j=1,\ldots,J-1$; the last interval has a length $d_J=10^{-J}$ to ensure $\sum_{j=1}^{J}\delta_j=1$. We next perform a $P$-level uniform partitioning on each interval, i.e., each interval is now partitioned by $P$ equally-spaced points. Finally, we invert the partitioned cdf to obtain a typical set $\mathcal{S}_n^\ell$ of cardinality~$\ell=J\times P$. Also, as the proposed partitioning is non-uniform, $\mathcal{S}_n^\ell$ needs to be associated a probability set: the probability of an element computed from the $j$th interval is $\delta_j=d_j/P$. It can be shown (see Section~\ref{results_shadowing_single}) that using $J=25$ intervals containing $P=900$~points each ---~which results in a typical set that contains only $\ell=25\times900=22,500$~elements \footnote{To produce moments of the same accuracy, the traditional uniform partitioning approach would require about $\ell=900\times10^{25}$ points.}~--- guarantees that up to third-order moments derived from the typical set are within 1\% of the exact values for all $\sigma_{\textnormal{dB}}$. 

\subsubsection{\label{N_interferers}Multiple interferers}

We now focus on finding an $\mathcal{L}$-element typical set ---~denoted~$\mathcal{S}^{\mathcal{L}}$~--- for the interference gain $G$ that must be computed from $N$~typical sets~$\mathcal{S}_n^\ell$, $n=1,2,\dots,N$.

We first note that interferer~$n$'s typical set can be directly obtained by weighting each element of $\mathcal{S}_n^\ell$ by its average path loss~$\lambda_n$; we will denote interferer~$n$'s typical set by~$\lambda_n\mathcal{S}_n^\ell$. Let us now find a way to produce the ensemble~$\mathcal{S}^{\mathcal{L}}$ from the typical sets~$\lambda_n\mathcal{S}_n^\ell$. 

Ideally, $\mathcal{S}^{\mathcal{L}}$ should be constructed by considering all combinations of the elements of the typical sets~$\lambda_n\mathcal{S}_n^\ell$, but the cardinality of the resulting set, $\mathcal{L}=\ell^N=\left(JP\right)^N$, would rapidly become prohibitive as the number~$N$ of interferers increases. 

To get rid of this complexity, we point out that the above-mentioned ideal (exhaustive) solution can also be viewed as an exhaustive combination of intervals ($J^N$ combinations) associated with an exhaustive combination of elements within each interval combination ($P^N$ combinations). And we observe that the most important part of this exhaustive solution pertains to the combination of \emph{intervals}, i.e., the combination of elements belonging to interval~$j$ of typical set $\lambda_n\mathcal{S}_{n}^\ell$ with elements belonging to interval~$k$, $k\neq j$, of typical set $\lambda_m\mathcal{S}_{m}^\ell$, $m\neq n$. 
So a way to construct a (near optimal) typical set for $G$ could be to perform exhaustive combinations of the intervals (as in the exhaustive solution), and to approximate the exhaustive combination of the elements within each interval combination by the following procedure: for each of the $J^N$~combinations of $N$ $P$-point intervals, 
\begin{itemize}
\item Perform a random permutation of the $P$ elements within each of the $N$ $P$-point intervals\footnote{Two interval combinations of the same rank~$j$ are supposed to be orthogonal because of the high number of points in each interval ($P=900$), which guarantees the independance of permutations.}; 
\item Add up these N permuted $P$-point intervals to obtain one resulting P-element interval. 
\end{itemize}
This last $P$-element interval approximates the $P^N$-element interval that would have resulted from an exhaustive combination of elements within the considered interval combination. Now, as there are $J^N$ interval combinations, the resulting typical set would contain $J^N P$~elements, which can still be prohibitive, so this second solution ---~which we will refer to as the \emph{near-optimal} solution~--- can not be applied as such. 

We eventually propose a novel approach which makes use of this near-optimal solution and is based on the following two-step algorithm:
\begin{description}[\IEEEsetlabelwidth{Step X)}]
\item[Step 1] Apply exhaustive combinations of intervals to a subset of $M$~interfering links;
\item[Step 2] Perform Monte Carlo simulations for the $N-M$ remaining links. 
\end{description}
We now detail the principle of the proposed method. In Step~1, we apply the near-optimal solution described above, but to a subset of $M<N$ interfering links which we will call \emph{compelled} links. The compelled links are chosen to have the highest average path losses ($\lambda_1\geq\cdots\lambda_M\geq\cdots\lambda_N$) so as to minimize errors in other (non-compelled) interfering links. The exhaustive combination of the $J$~intervals for $M$~compelled links obtained from the near-optimal solution thus results in \emph{one} set of $J^M P$ elements. In Step~2, we build up a $J^M P$-element set for \emph{each} of the $N-M$ remaining, non-compelled, links by performing $J^M$ random drawings of intervals according to the probability set $\left\{\delta_j\right\}$, $j=1,2,\dots,J$. As in the near-optimal solution, a random permutation of the elements is applied at each drawing. The ensemble of amplitudes of the intercell interference gain $G$ ---~the so-called typical set $\mathcal{S}^{\mathcal{L}}$~---is then constructed by adding up these $N-M+1$ sets; it is of cardinality $\mathcal{L}=J^M P$. Associated to $\mathcal{S}^{\mathcal{L}}$ is a probability set determined as follows: to each interval is associated a \emph{weight} which is the product of probabilities $\delta_k$ of intervals issued from compelled links (for non-compelled links, probabilities are accounted for by means of the random selection process); these weights are then normalized to obtain probabilities. Finally, the histogram of the interference gain $G$ can be constructed from these resulting amplitude and probabiliy sets. It is important to note, however, that, as a random drawing process is involved, a number of iterations might be needed in order for this process to converge (elements of $\mathcal{S}^{\mathcal{L}}$ and associated probabilities are averaged at each iteration). We will call this semi-analytical technique the \emph{Monte Carlo-panel method} (MCP, in short)\footnote{The term 'panel' refers to survey panels used by polling organizations.}. 

\begin{figure}[t]
\begin{center}
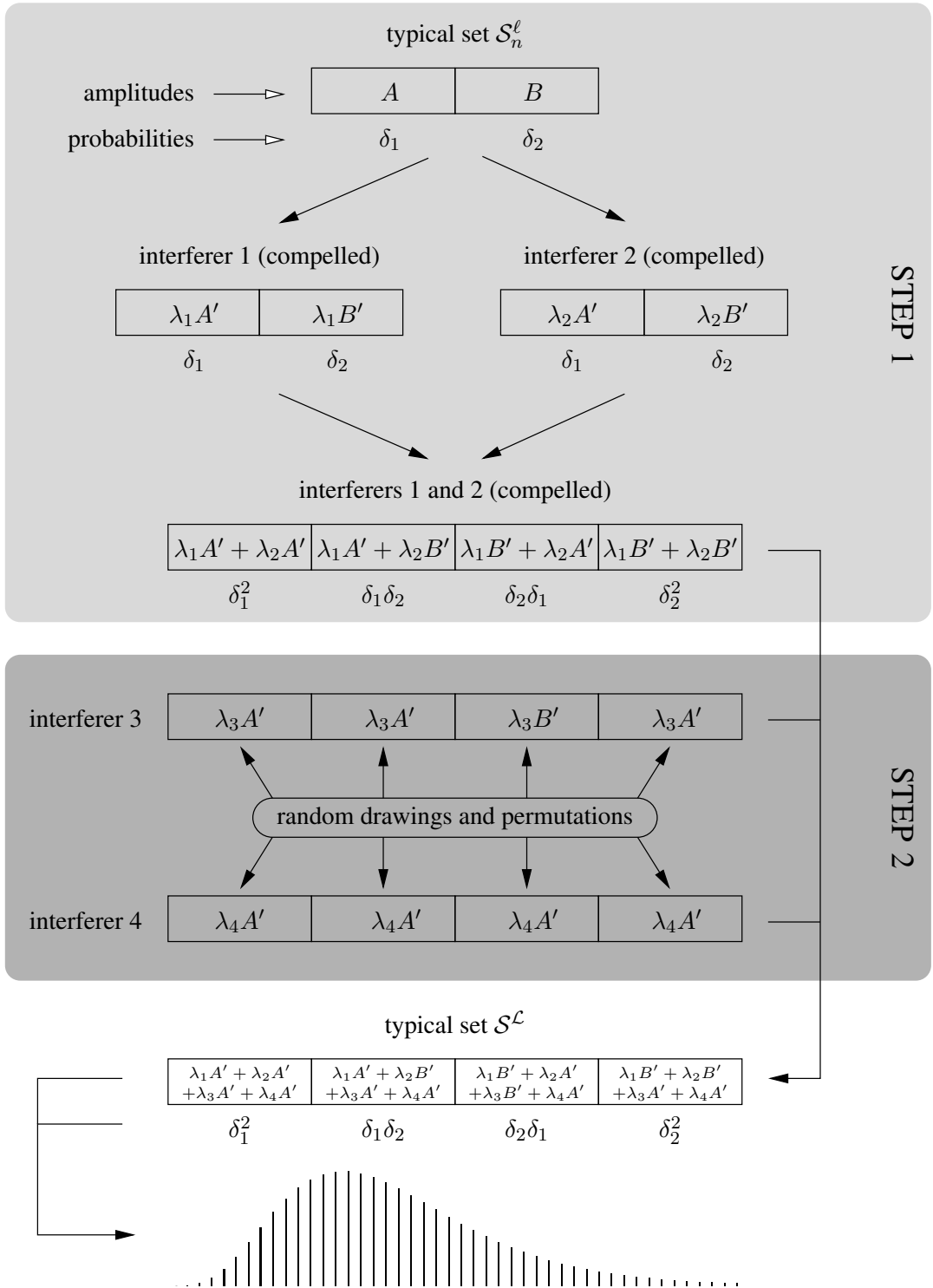
\end{center}
\caption{\label{illustration_MCP}Illustration of the MCP method for $N=4$ interfering cells, $M=2$ compelled links, and $J=2$ intervals per link (denoted~$A$ and~$B$, with respective probabilities~$\delta_1$ and~$\delta_2$). Each $A'$ (resp.~$B'$) represents \emph{one} random permutation of $A$ (resp.~$B$).}
\end{figure}

The MCP method is illustrated on Fig.~\ref{illustration_MCP} for $N=4$ interfering cells, $M=2$ compelled links, and $J=2$ intervals per typical set (these intervals ---~denoted $A$ and $B$~--- have probabilities $\delta_1=0.9$ and $\delta_2=0.1$ respectively, and each one of them contains $P$ elements). Step~1 of the alogrithm is summarized in the light-grey shaded box: intervals from typical sets $\mathcal{S}_1^\ell$ and $\mathcal{S}_2^\ell$ (corresponding to compelled interfering links 1 and 2, and weighted by their respective average path losses $\lambda_1$ and $\lambda_2$) are combined together, as described in the near-optimal solution, to obtain a set of amplitudes of cardinality~$4P$ representative of the two compelled links; associated to this set of amplitudes is a set of weights $\left\{0.81,0.09,0.09,0.01 \right\}$. The dark-grey shaded box summarizes Step~2: for each non-compelled interfering link, a $4P$-element set of amplitudes is made up by 4 intervals ($A$ or $B$) drawn according to the probability set $\left\{0.9,0.1 \right\}$ and applied random permutations. The typical set $\mathcal{S}^{\mathcal{L}}$ (with $\mathcal{L}=4P$ in our example) is then obtained by summing up together all these sets. The histogram of the interference gain $G$ is constructed from $\mathcal{S}^{4P}$ and the associated probability set\footnote{The probability set is obtained by normalizing the set of weigths.}. Note that \emph{one} random permutation of the interval (permuted intervals have been assigned the prime symbol) is performed at \emph{each} (compelled or random) manipulation of an interval. 

Implementing the MCP method however requires cautiousness. In non-compelled links, random drawings of intervals are performed based on the probability set $\left\{\delta_j\right\}$, $j=1,2,\dots,J$. In this process, lowest-probability intervals, which contain the highest interference gains, are totally ignored for two reasons. The first reason pertains to the fact that obtaining a significant frequency of appearance of such rare events would require a prohibitive number of simulation runs. The second reason is due to limitations inherent to software simulation tools which use pseudo-random number generators to generate sequences of 'random' numbers belonging to a fixed set of values. In order to take into account the ignorance of the contribution of the highest interference gains of the $N-M$ non-compelled interfering links in the probability set~$\left\{\delta_j\right\}$, we suggest the following workaround: in these links, we intentionally make exclusive use of the $\mathcal{J}$, $1\leq\mathcal{J}<J$, first intervals, and we associate them a \emph{loaded} probability set $\left\{\delta'_j\right\}$ defined as follows:
\begin{equation}
\delta_{j}'=
\begin{cases}
\alpha\delta_{j} & \textnormal{for }1\leq j\leq\mathcal{J}\\
0 & \textnormal{for }\mathcal{J}+1\leq j\leq J
\end{cases}
\label{deltaj_alpha_deltaj_0}
\end{equation}
where 
\begin{equation}
\alpha=\frac{1}{\underset{j=1}{\overset{\mathcal{J}}{\sum}}\delta_{j}}
\label{alpha_1_sum_j1_J_deltaj}
\end{equation}
is a normalizing constant such that $\sum_{j=1}^{\mathcal{J}}\delta_{j}'=1$ (using the particular non-uniform partitioning described previously, we have: $\alpha=1/\left(1-0.1^{\mathcal{J}}\right)\gtrsim1$). 

Now, as was mentioned before, high amplitudes play an important role in terms of moments. Although the impact of neglecting them in non-compelled links is globally limited because these links are weighted by smaller average path losses~$\lambda_n$ ($n=M+1,\dots,N$), it has to be compensated in order to satisfy the $1$st-order moment constraint (i.e., the sampled mean has to converge to the exact value\footnote{We recall that the mean $\mathbb{E}\left\{G\right\}$ is of particular importance because it is proportional to the average interference power.}). For this purpose, small (resp. large) amplitudes need to be underweighted (resp. overweighted). Thus, an underweighting multiplicative factor, denoted $f^-$, is applied to amplitudes of the $\mathcal{J}$ first intervals of compelled links; similarly, an overweighting multiplicative factor $f^+$ is applied to amplitudes of the last $N-\mathcal{J}$ intervals. (Computation details of factors $f^-$ and $f^+$ are given in Appendix~\ref{appendix3}.) 

Let us last notice that the choice for values of $M$ and $\mathcal{J}$ is a trade-off between differents aspects: cardinality of the resulting typical set (i.e., tractable number of points), number of simulation runs and accuracy of the histogram. We have determined that $M=2$ and $\mathcal{J}=3$ meet all these requirements.

\section{\label{results}Numerical results}
In this section, we present numerical results related to the different methods introduced in the preceding section. In Section~\ref{results_fading}, we first examine the validity of the original approximation introduced in Section~\ref{methodology}, stating that the interference gain~$G_n$ (and, consequently, $G$) does not depend on the user's position within its cell. For this purpose, we compare the approximation of $G$ given by~\eqref{Gn_lambdan_Gfn} with the 'exact' formula~\eqref{Gnrn_Gplnrn_Gfn}. Then, in Section~\ref{results_shadowing_single}, we obtain the histogram of the interference gain~$G_n$ (one single interferer) by applying the non-uniform partitioning generalized inverse method described in~\ref{one_interferer}. Finally, the MCP method (see~\ref{N_interferers}) is used to build up the histogram of the interference gain for multiple interferers in Section~\ref{results_shadowing_multiple}. 

We use the following simulation parameters. We consider a system functioning at 1 GHz. We fix the cell radius to $R=700$ m, $d_{0}=10$ m, and the pathloss exponent to $3.2$, which corresponds to a typical urban environment, as described in the COST-231 reference model \cite{COST231}. The reference distance is chosen to be equal to $2R$. Average path losses $\lambda_{n}$, $n=1,2,\dots,N$, are determined numerically using \eqref{lambdan_Gpln_Er} and are summarized in Table~\ref{tab_lambdan}.  

\subsection{\label{results_fading}No shadowing}

\begin{figure}[t]
\begin{center}
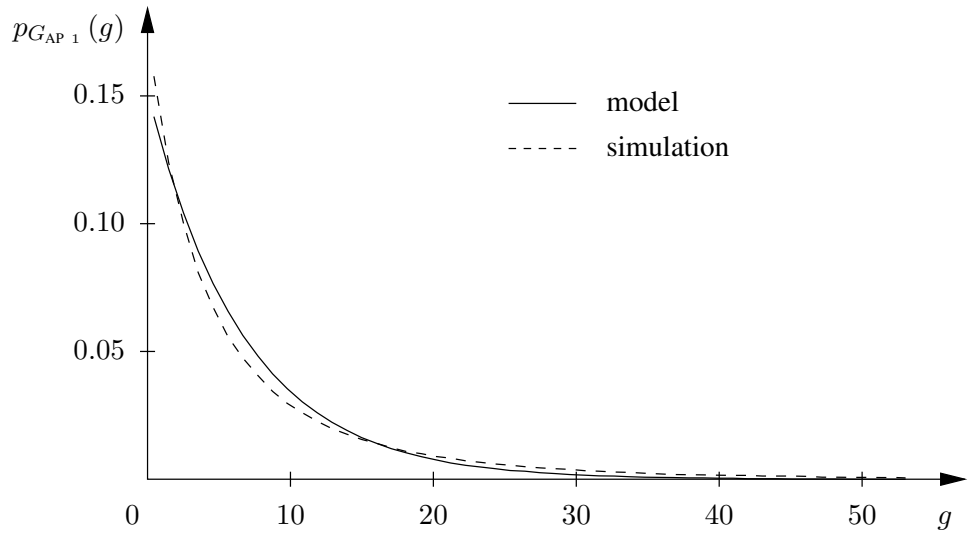
\end{center}
\caption{\label{fig_pdf_AP1}Simulated vs. modeled pdf of the intercell interference power with no shadowing when AP~$1$ is the only interferer. Since AP~$1$ produces the largest dynamics for the interference power undergone by a user in the grey-shaded sector of Fig. \ref{network_symmetry} with only one interfering cell, these curves correspond to the worst-case scenario for validating our approximation.}
\end{figure}

\begin{figure}[t]
\begin{center}
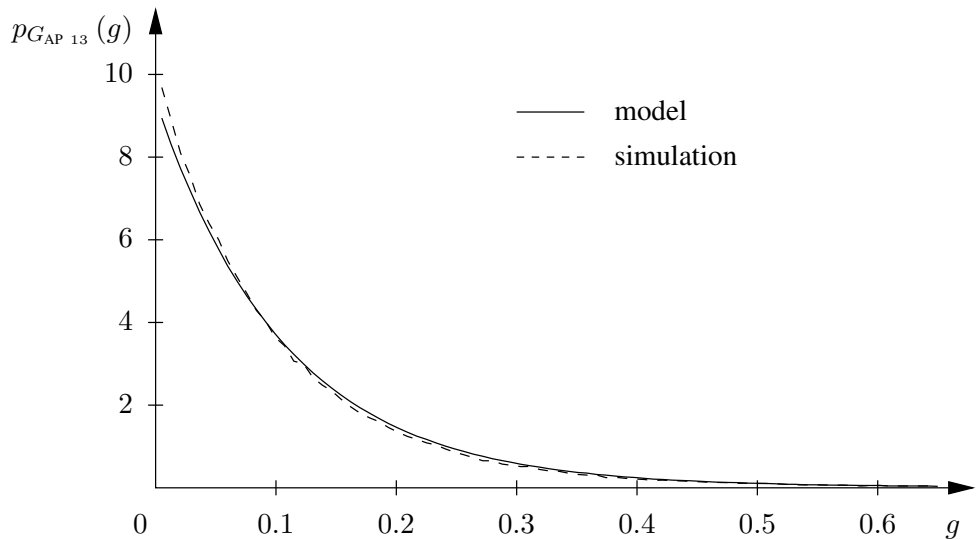
\end{center}
\caption{\label{fig_pdf_AP13}Simulated vs. modeled pdf of the intercell interference power with no shadowing when AP~$13$ is the only interferer. AP~$13$ produces the smallest dynamics for the interference power undergone by a user in the grey-shaded sector of Fig.~\ref{network_symmetry} with only one interfering cell (best match for our model).}
\end{figure}

In this section, we evaluate the proposed approximation \eqref{Gn_lambdan_Gfn} against Monte Carlo simulations performed on \eqref{Gnrn_Gplnrn_Gfn}. We first consider the contribution of one interfering cell and, in this regard, we examine two opposite scenarios: one for which the investigated interferer (i.e., AP~1) produces the largest dynamic range for the intercell interference power undergone by a user in the grey-shaded triangular area of Fig.~\ref{network_symmetry}; the other one for which the investigated cell (i.e., AP~13) has the smallest dynamics. Obviously, both dynamics differently impact the accuracy of our model. Note that, in both cases, the sum of interference gains \eqref{G_sum_lambdan_Gfn} reduces to one exponential r.v. Modeled and simulated pdf's for above-mentioned cases (a) and (b) are plotted in Fig.~\ref{fig_pdf_AP1} and Fig.~\ref{fig_pdf_AP13} respectively, and the good match of the curves shows that the proposed method is a good approximation. 

\begin{figure}[t]
\begin{center}
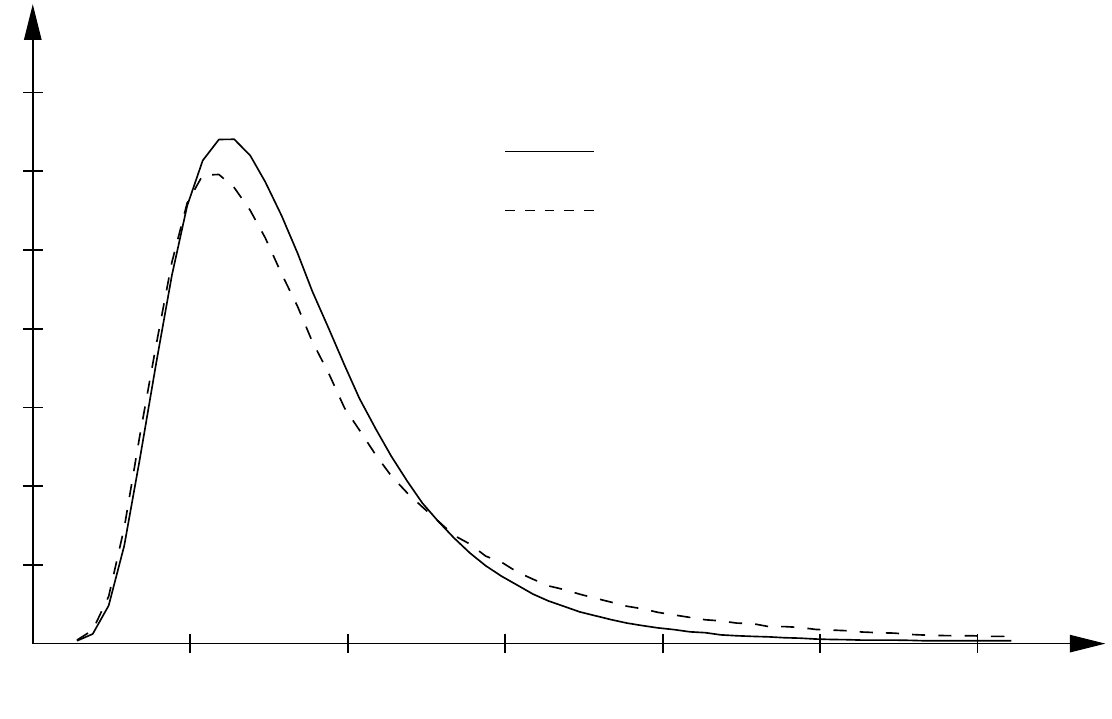
\end{center}
\caption{\label{fig_pdf_FR1}Simulated vs. modeled pdf of the intercell interference power~$G$ for frequency reuse pattern FR1.}
\end{figure}

\begin{figure}[t]
\begin{center}
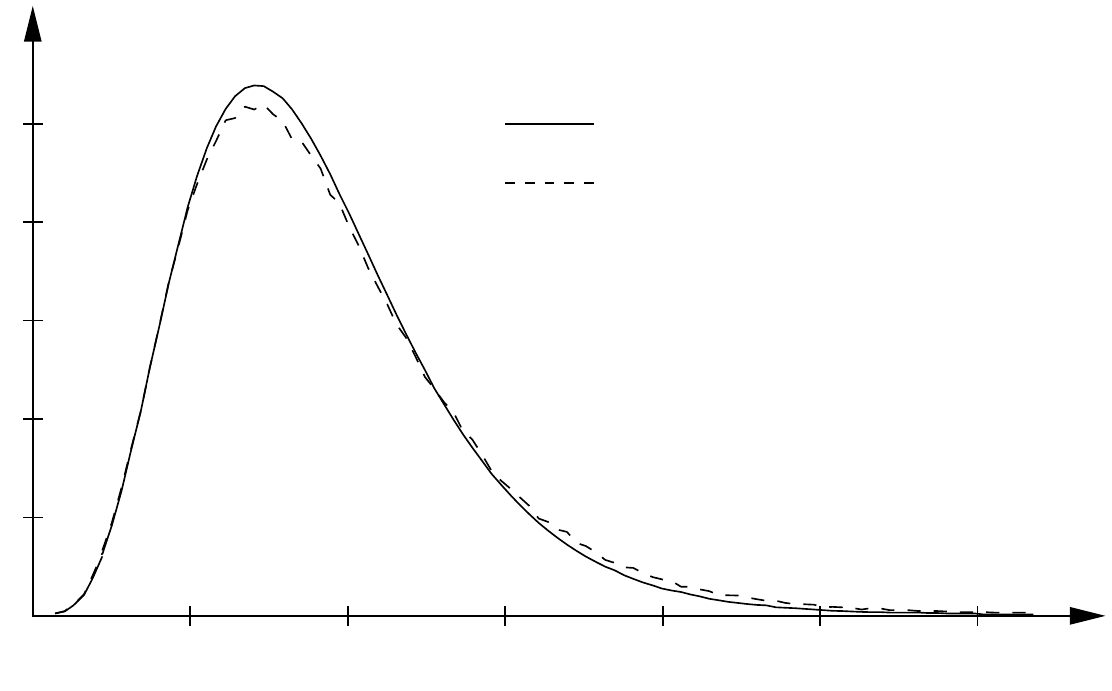
\end{center}
\caption{\label{fig_pdf_FR3}Simulated vs. modeled pdf of the intercell interference power~$G$ for frequency reuse pattern FR3.}
\end{figure}

We then consider the whole set of interfering cells ($N$~interferers) under frequency reuse patterns FR1 and then FR3, for which results are shown in Fig.~\ref{fig_pdf_FR1} and Fig.~\ref{fig_pdf_FR3} respectively. We see that simulated and modeled probability laws \eqref{G_sumn1N_Gplnrn_Gfn_Gsn} and~\eqref{G_sum_lambdan_Gfn} respectively closely match for both frequency reuse patterns. We also note that simulated and approximated curves are closer to one another for FR3 than they are for FR1. As explained before for the single-interferer scenario, fluctuations of actual pathlosses $G_{\textnormal{pl},n}\left(r_{n}\right)$, $n=7,...,18$, can be assumed to have about the same dynamic range, but these dynamics are smaller than those of gains $G_{\textnormal{pl},n}\left(r_{n}\right)$, $n=1,...,6$. 

\subsection{\label{results_shadowing_single}Shadowing, one interferer}

In this section, we make use of the non-uniform partitioning generalized inversion method introduced in Section~\ref{one_interferer} to obtain a typical set for the interference gain of one interferer. Table~\ref{moments_single} presents the three first moments computed from typical set $\mathcal{S}_n^\ell$, as compared with the exact moments of the distribution of the interference gain $G_n$. We see that moments issued from the typical set are far beyond the 1\% accuracy requirement. The proposed method also outperforms the Monte Carlo simulation technique, which cannot be guaranteed to converge for such a small number of points. 

\begin{figure}[t]
\begin{center}
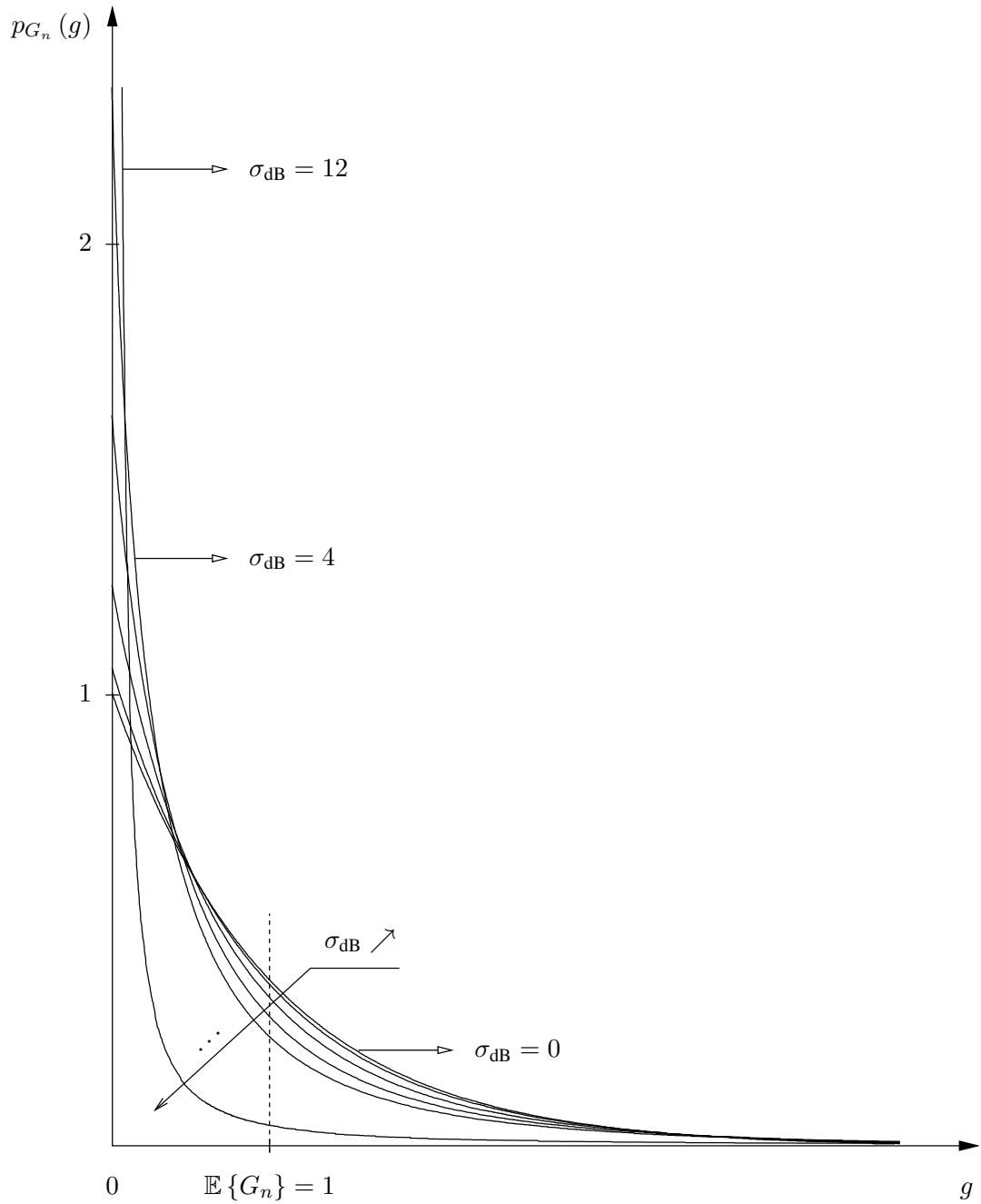
\end{center}
\caption{\label{fig_histogram_single}Histograms of the interference gain $G_n$ (one interferer) for different values of $\sigma_{\textnormal{dB}}$.}
\end{figure}

Histograms of the interference gain $G_n$ computed from typical set $\mathcal{S}_n^\ell$ is illustrated on Fig.~\ref{fig_histogram_single} for different values of $\sigma_{\textnormal{dB}}$. 

\subsection{\label{results_shadowing_multiple}Shadowing, multiple interferers}

We now evaluate the MCP method developed in Section \ref{N_interferers}. We have determined that $20,000$ iterations of the base MCP algorithm guarantee that the $1$st-order moment computed from any typical set (whatever $\sigma_{\textnormal{dB}}$ value is considered) converges to its exact value~\eqref{nu1}. Table~\ref{moments_single} presents the values of the $1$st-order moment of $G$, both exact (analytical) and approximated (computed from the typical set). We can see that the proposed method performs very well for the whole range of $\sigma_{\textnormal{dB}}$ values. 

\begin{figure}[t]
\begin{center}
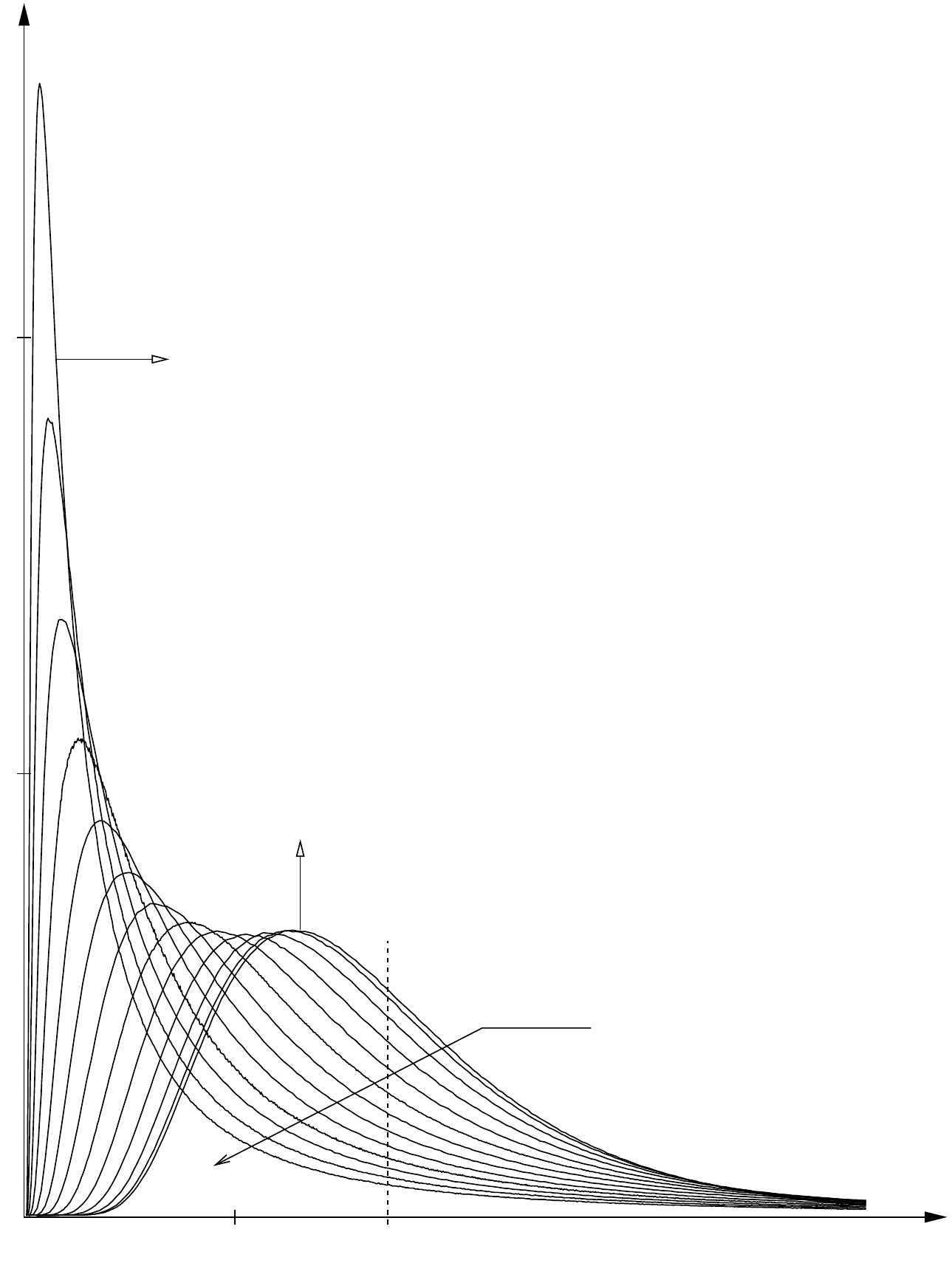
\end{center}
\caption{\label{fig_histogram_FR1}Histograms of the interference gain $G$ obtained by the MCP method (FR1 scenario).}
\end{figure}

\begin{figure}[t]
\begin{center}
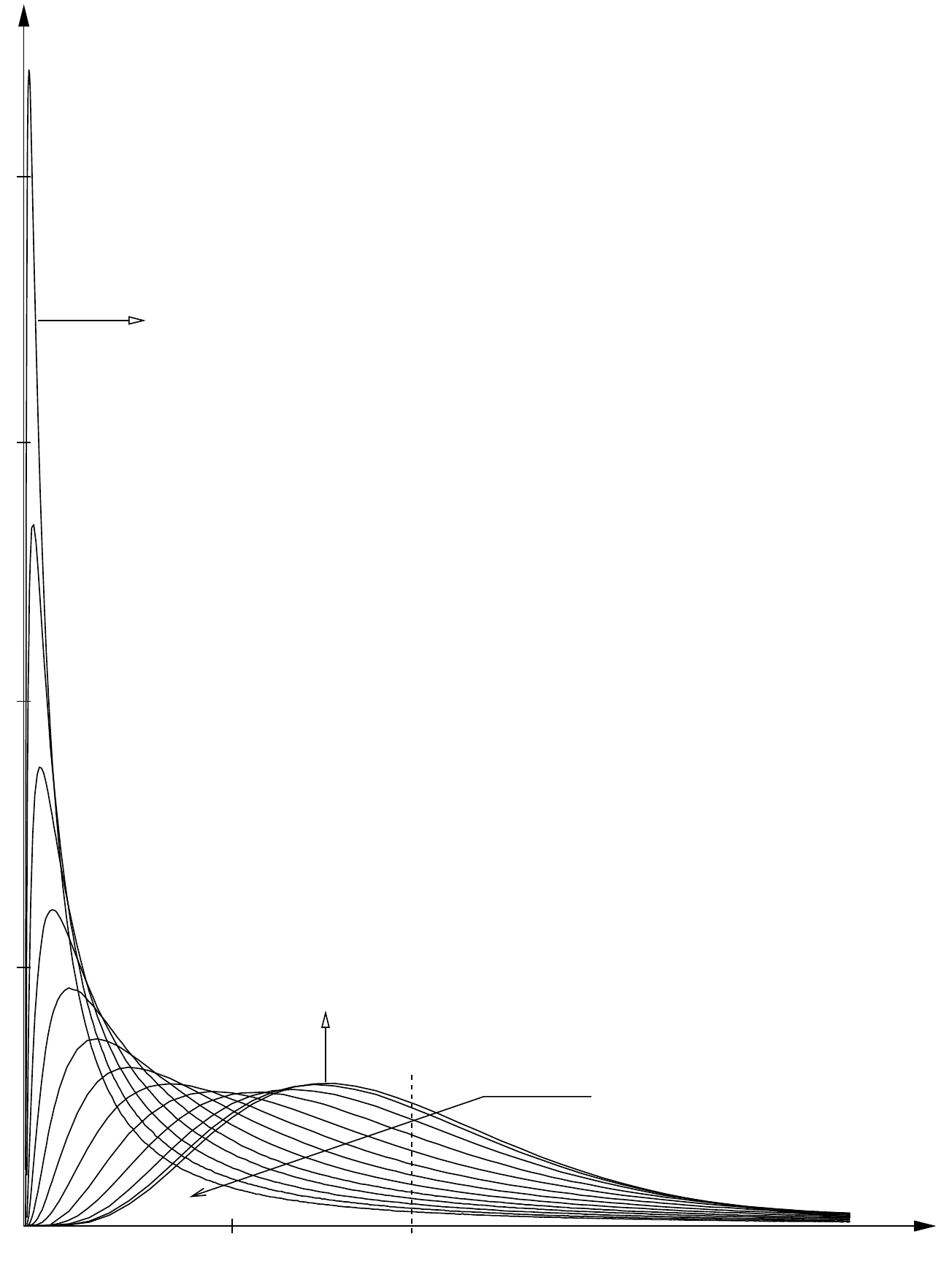
\end{center}
\caption{\label{fig_histogram_FR3}Histograms of the interference gain $G$ obtained by the MCP method (FR3 scenario).}
\end{figure}

Histograms of the interference gain $G$ computed from typical sets obtained by the MCP method are illustrated on Fig.~\ref{fig_histogram_FR1} (FR1 scenario) and Fig.~\ref{fig_histogram_FR3} (FR3 scenario) for different values of $\sigma_{\textnormal{dB}}$.

\section{\label{model}Statistical model}

In Section~\ref{methodology}, we developed analytical and numerical methods to build up a good approximation of the histogram of the interference gain~$G$. In this section, we aim at using this result to elaborate a statistical model for~$G$, i.e., a closed form expression of the probability law, characterized by the shadowing parameter~$\sigma_{\textnormal{dB}}$. This task is challenging in that one single parametric law is required, that is valid for propagation environments which considerably vary depending upon the shadowing phenomenon (parameter~$\sigma_{\textnormal{dB}}$), and that is applicable to various frequency reuse scenarios (FR1 and FR3). 

We initialize the modeling process by extracting usefull information from a carefull analysis of the histograms of the interference gain~$G$ (see Fig.'s~\ref{fig_histogram_FR1} and~\ref{fig_histogram_FR3}). We first note that $G$~is a positive continuous r.v. We then observe that all curves are asymmetric, and this property is even more pronounced for large values of~$\sigma_{\textnormal{dB}}$. In this case, G's pdf's also have a sharper peak and a longer, fatter tail, the last of which being a characteristic of heavy-tailed distributions (a.k.a. \emph{power distributions}), as already mentioned. 

Due to the strongly skewed nature of the interference gain distribution for large~$\sigma_{\textnormal{dB}}$'s, a power-type statistical model turns out to be suitable here. In this regard, a Pareto-like distribution seems to be a good candidate, so we focus, in first approximation, on a 3-parameter Burr-type \textsc{xii} distribution~\cite{Johnson1994}. The Burr distribution has a flexible shape and controllable location and scale, which makes it appealing to fit any given set of unimodal data that exhibits a heavy-tail behavior (e.g., it is an appropriate model for characterizing insurance claim sizes). However, as 3 parameters seem to not be sufficient to correctly characterize the interference gain distribution under those particularly tight constraints, another law is required, which offers greater flexibility to match the whole range of $\sigma_{\textnormal{dB}}$ values. Such a flexibility is provided by introducing an additional shape parameter into the Burr distribution, based on the following property \cite{Gupta2009}: if $F(x)$ is a cdf, so is $\left(F(x)\right)^\eta$, $\forall\eta>0$. Thus, we have established a new Burr-based probability law, whose cdf ---~denoted~$F_G\left(x\right)$~--- is given by
\begin{equation}
F_{\textnormal{G}}\left(x\right)=\left(1-\dfrac{1}{\left(1+\left(\dfrac{x}{\beta}\right)^{\alpha}\right)^{k}}\right)^{\eta},\qquad\begin{cases}
x>0\\
\eta>1\\
\alpha,k,\beta>0\end{cases}
\label{FGx}
\end{equation}
where $\eta$, $\alpha$ and $k$ are the shape parameters, and $\beta$ is the scale parameter of the distribution. G's pdf ---~denoted~$p_G\left(x\right)$~--- can be easily obtained by deriving~\eqref{FGx}: 
\begin{equation}
p_G\left(x\right)=\dfrac{\eta\alpha k}{\beta}\left(\dfrac{x}{\beta}\right)^{\alpha-1}\dfrac{\left(\left(1+\left(\dfrac{x}{\beta}\right)^{\alpha}\right)^{k}-1\right)^{\eta-1}}{\left(1+\left(\dfrac{x}{\beta}\right)^{\alpha}\right)^{k\eta+1}}.
\label{pGx}
\end{equation}

\begin{figure}[t]
\begin{center}
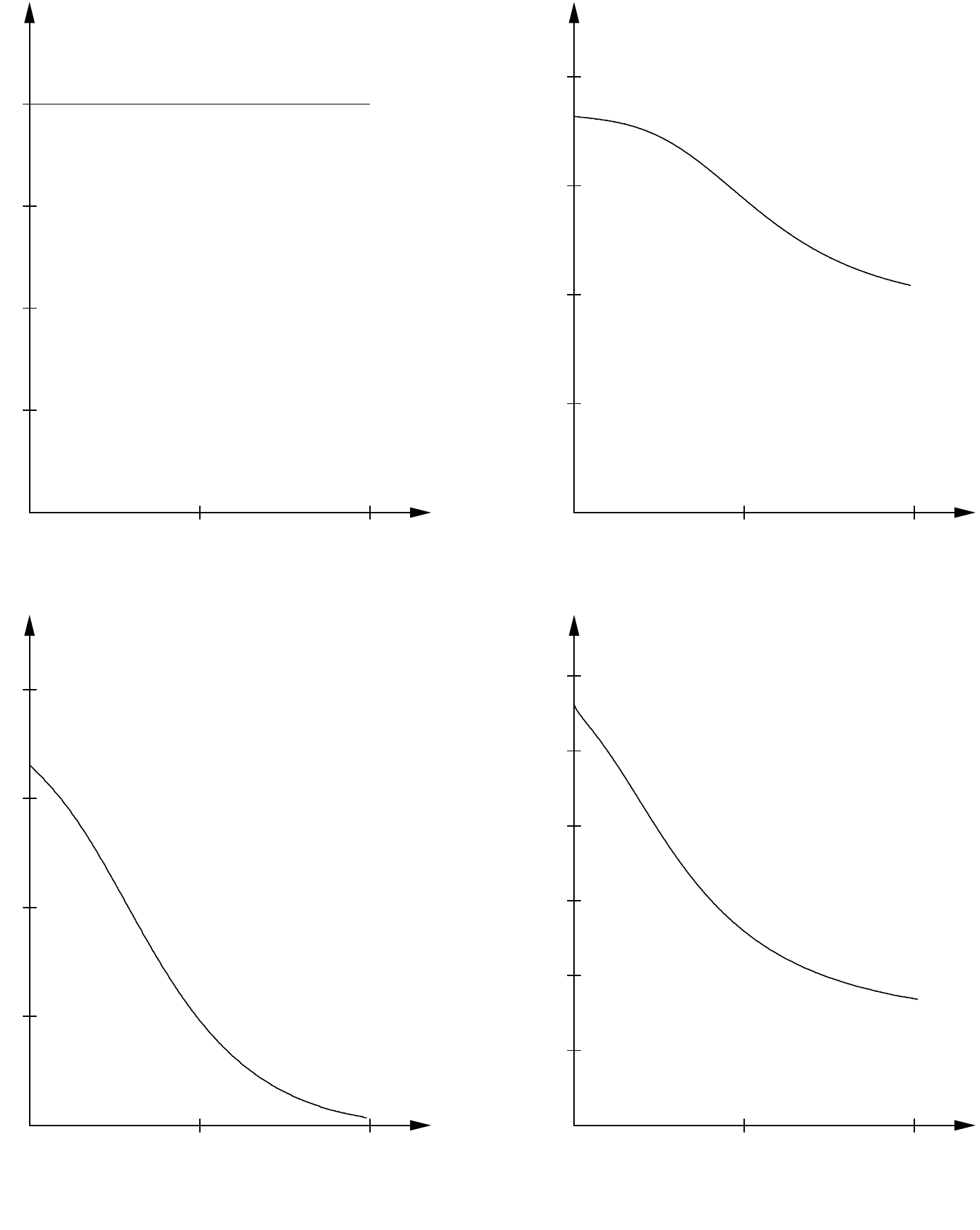
\end{center}
\caption{\label{fig_f_FR1}Empirical laws~$\eta$, $\alpha$, $k$, and~$\beta$ as functions of $\sigma_{\textnormal{dB}}$ (FR1 scenario).}
\end{figure}

\begin{figure}[t]
\begin{center}
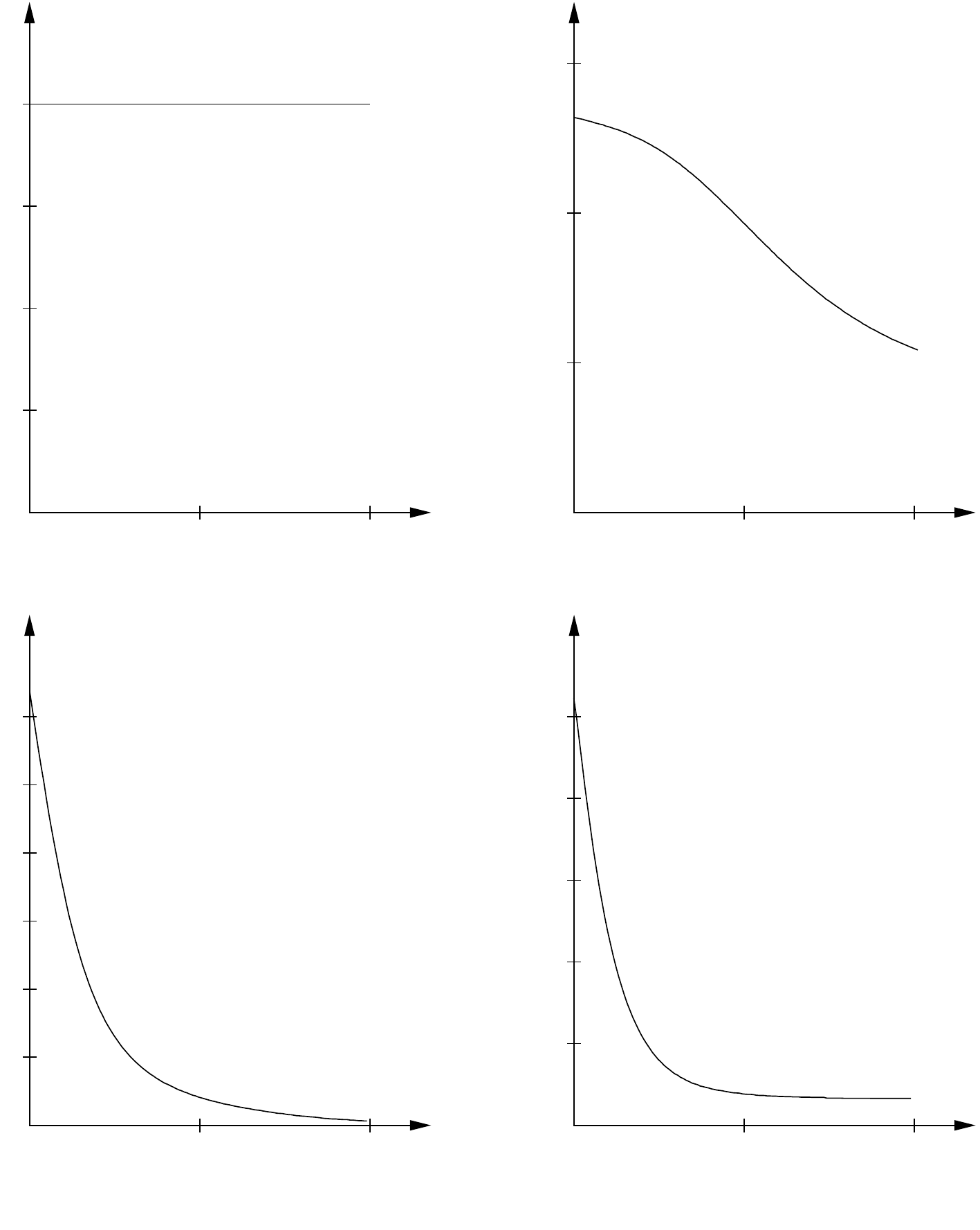
\end{center}
\caption{\label{fig_f_FR3}Empirical laws~$\eta$, $\alpha$, $k$, and~$\beta$ as functions of $\sigma_{\textnormal{dB}}$ (FR3 scenario).}
\end{figure}

\begin{figure}[t]
\begin{center}
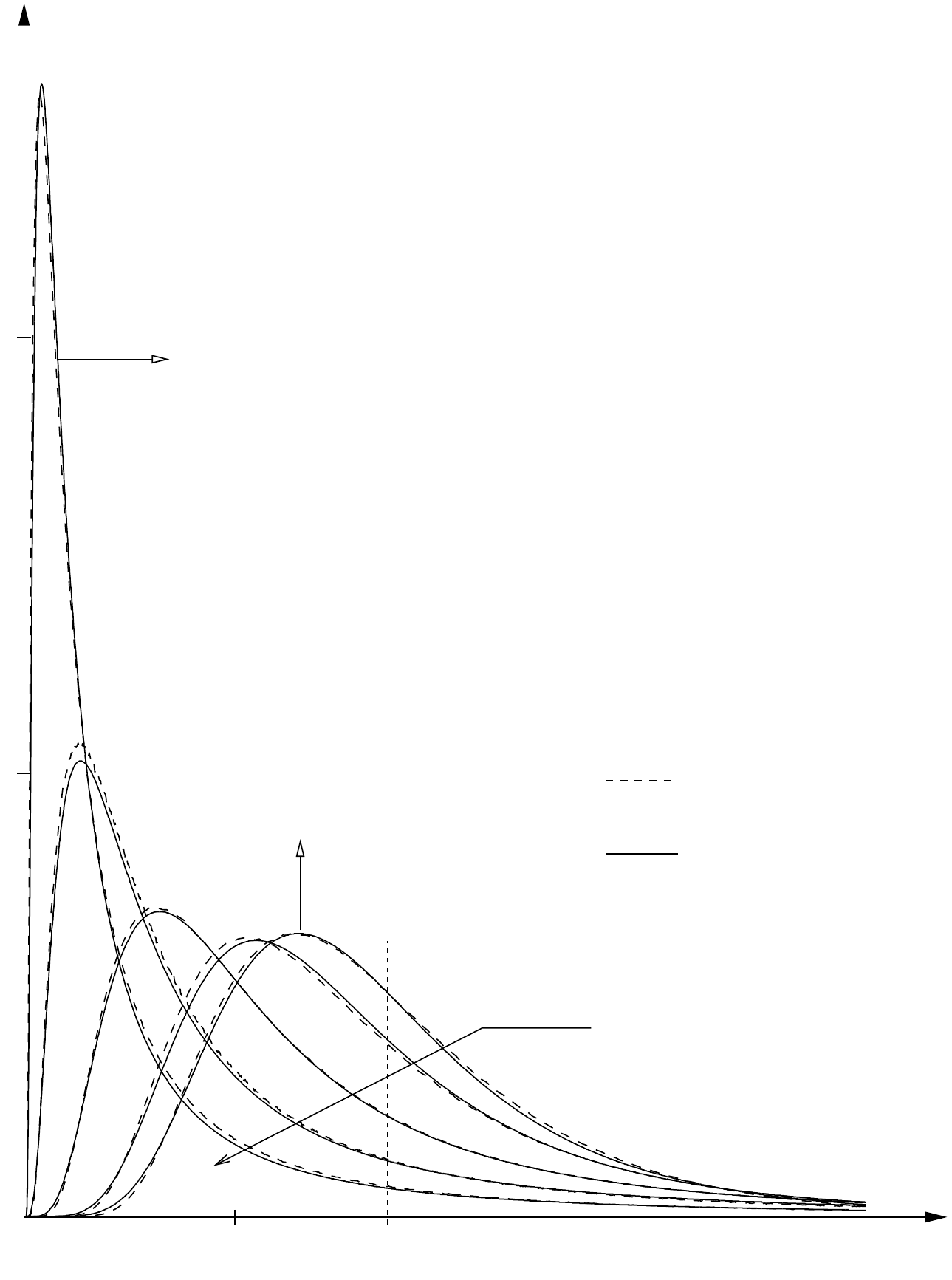
\end{center}
\caption{\label{fig_histogram_Burr_FR1}Comparison of MCP histograms and modeled cdf of the interference gain~$G$ for $\sigma_{\textnormal{dB}}=0,3,6,9,12$ (FR1 scenario).}
\end{figure}

\begin{figure}[t]
\begin{center}
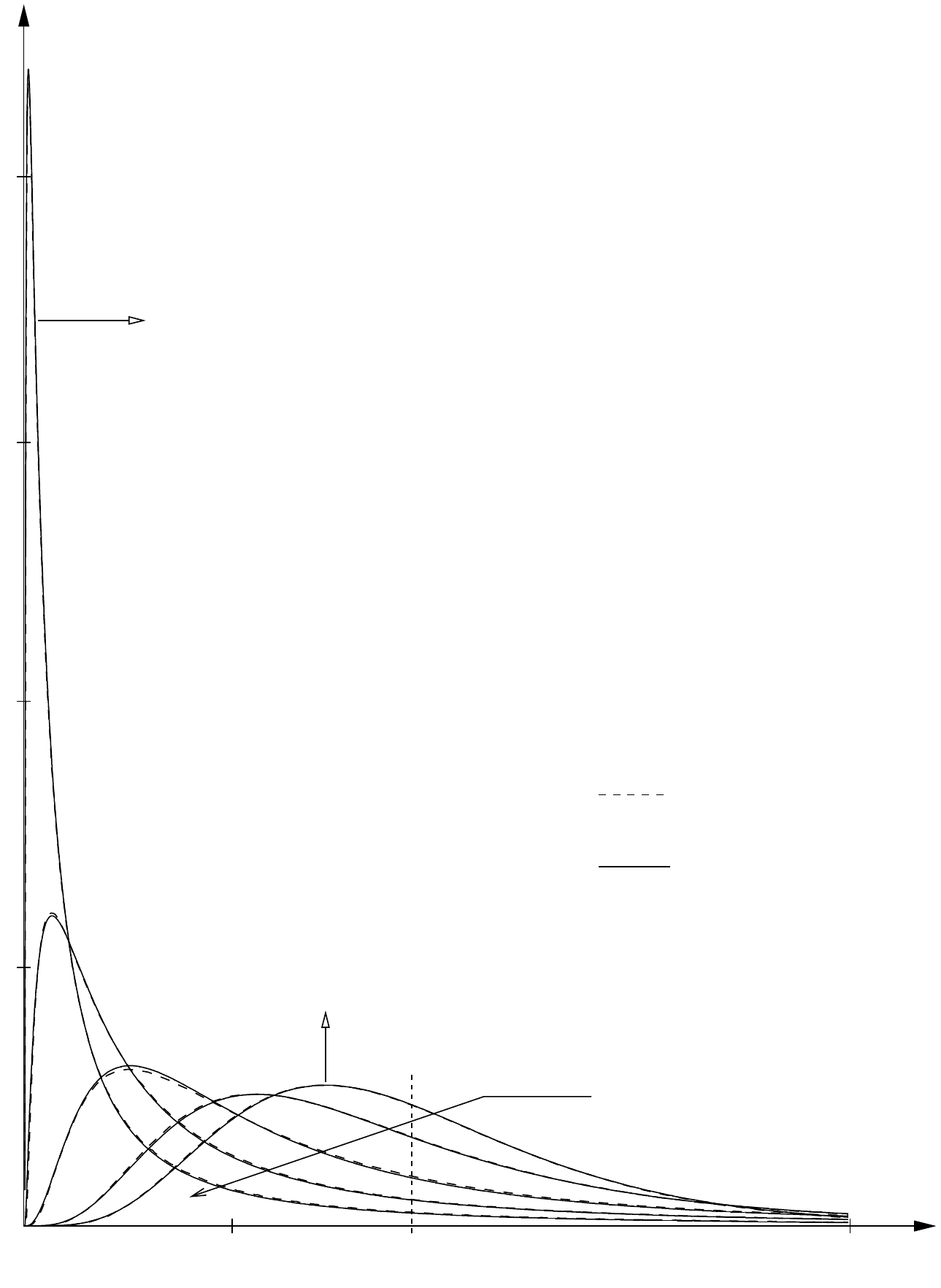
\end{center}
\caption{\label{fig_histogram_Burr_FR3}Comparison of MCP histograms and modeled cdf of the interference gain~$G$ for $\sigma_{\textnormal{dB}}=0,3,6,9,12$ (FR3 scenario).}
\end{figure}

\begin{figure}[t]
\begin{center}
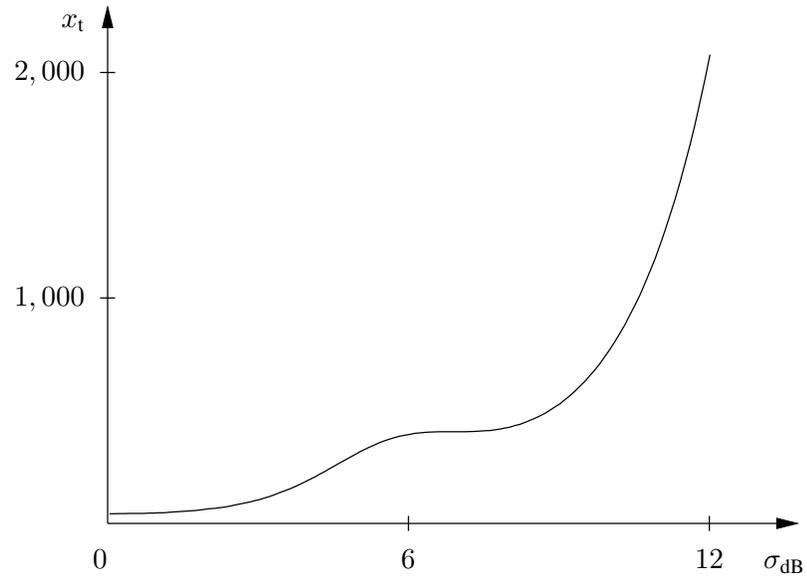
\end{center}
\caption{\label{fig_xt_FR1}Truncation gain $x_{\textnormal{t}}$ as a function of~$\sigma_{\textnormal{dB}}$ (FR1 scenario).}
\end{figure}

\begin{figure}[t]
\begin{center}
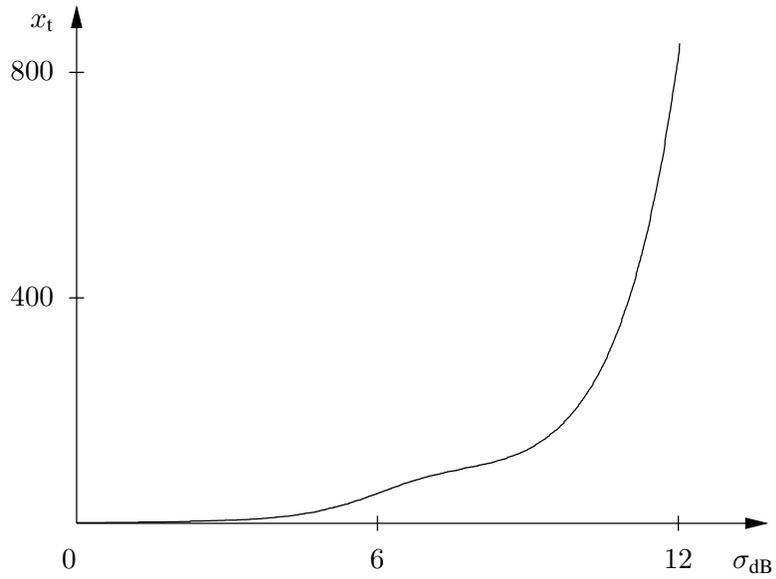
\end{center}
\caption{\label{fig_xt_FR3}Truncation gain $x_{\textnormal{t}}$ as a function of~$\sigma_{\textnormal{dB}}$ (FR3 scenario).}
\end{figure}

We next establish a parametric family of functions (parameterized by~$\sigma_{\textnormal{dB}}$) for the interference gain~$G$ by determining empirical formulas for parameters $\eta$, $\alpha$, $k$, and~$\beta$. For this purpose, we propose that all parameters (whatever frequency scenario is considered) be modeled by the same 6-parameter function~$f$ that has the following expression:
\begin{equation}
f\left(\sigma_{\textnormal{dB}}\right)=a_1 +a_2\cdot\dfrac{1-\dfrac{\sigma_{\textnormal{dB}}}{a_3}}{\left(1+\left(\dfrac{\sigma_{\textnormal{dB}}}{a_3}\right)^{a_4}\right)^{\frac{1}{a_4}}}\cdot\dfrac{1}{1+\left(\dfrac{\sigma_{\textnormal{dB}}}{a_5}\right)^{a_6}},
\label{}
\end{equation}
where coefficients $a_i$, $i=1,2,\dots,6$ have been determined empirically and are summarized in Table~\ref{parameters_e_a_k_b}. Corresponding empirical laws $f$, as functions of $\sigma_{\textnormal{dB}}$, are plotted on Fig.~\ref{fig_f_FR1} (FR1 scenario) and Fig.~\ref{fig_f_FR3} (FR3 scenario). The pdf's of the proposed statistical model are superimposed on histograms obtained by the MCP method for different values of the shadowing parameter~$\sigma_{\textnormal{dB}}$ on Fig.~\ref{fig_histogram_Burr_FR1} (resp. Fig.~\ref{fig_histogram_Burr_FR3}) for the FR1 (resp. FR3) scenario. 
We now come to the last step of our modeling process. As seen earlier, MCP-obtained histograms and the proposed Burr-based distributions closely match for the whole range of~$\sigma_{\textnormal{dB}}$. However, care must be taken in defining the range of gains for which our model is valid. And indeed, the Burr-based statistical law needs to be truncated at a maximum value ---~denoted~$x_{\textnormal{t}}$~--- defined in such a way that the $1$st-order moment constraint holds, which we can write
\[
\int_{0}^{x_{\textnormal{t}}}x p_G(x)\textnormal{d}x=\mathbb{E}\left\{G\right\},
\]
where $\mathbb{E}\left\{G\right\}$ is the exact mean~\eqref{nu1}. As a consequence of this truncation process, a normalizing factor, 
\begin{equation}
A=\dfrac{1}{1-\mathbb{P}\left(x>x_{\textnormal{t}}\right)},
\label{normalizing_factor}
\end{equation}
has to be incorporated in both the cdf and pdf of the elaborated model, which are then written $AF_G(x)$ and $Ap_G(x)$ respectively. Regarding the empirical law~$x_{\textnormal{t}}$ as a function of~$\sigma_{\textnormal{dB}}$, we also propose the same 5-parameter function for both FR1 and FR3 scenarios: 
\begin{equation}
x_{\textnormal{t}}\left(\sigma_{\textnormal{dB}}\right) =a_1\cdot\exp\left(\left(\dfrac{\sigma_{\textnormal{dB}}}{a_2}\right)^{a_3}\right) \cdot\exp\left(\exp\left(-\left(\dfrac{\sigma_{\textnormal{dB}}-a_4}{a_5}\right)^{2}\right)\right),
\label{}
\end{equation}
where coefficients $a_i$, $i=1,2,\dots,6$ have been determined empirically and are summarized in Table~\ref{parameter_xt}. Empirical laws $x_{\textnormal{t}}$, as functions of $\sigma_{\textnormal{dB}}$, are plotted on Fig.~\ref{fig_xt_FR1} (FR1 scenario) and Fig.~\ref{fig_xt_FR3} (FR3 scenario). The normalizing factor may be easily computed by replacing~$x_{\textnormal{t}}$ by its actual value in~\eqref{normalizing_factor}. 

\section{\label{Conclusion}Conclusion and Future work}
In this paper, we have proposed a methodology to estimate the statistics of the intercell interference power in the downlink of a multicellular network. In a propagation environment subject only to path loss and multipath Rayleigh fading, we have established an accurate approximated analytical expression for the interference power distribution. Then, considering the combined effects of path loss, lognormal shadowing and Rayleigh fading, we have proposed a semi-analytical method for the estimation of the pdf of the interference power. Finally, we have developed a statistical model parameterized by the shadowing parameter~$\sigma_{\textnormal{dB}}$ and valid on a large range of values ($\left[0,12\right]$~dB). It is our hope that the methods described in this paper are sufficiently detailed to enable the reader to apply them to other types of environments. 

A future work will pertain to improving the statistical interference power model by more closely linking the proposed model developed for a combined Rayleigh fading--lognormal shadowing environment to the 'exact' analytical formula obtained in the case where only Rayleigh fading was considered. Another perspective is to apply the proposed methods to other wireless network topologies (e.g., ad hoc networks,...).

\begin{appendix}

\subsection{\label{appendix0}Normalized channel power gain}
In this paper, we concentrate on the \emph{channel power gain} $H_n\left(r_n\right)=\left|h_n\left(m\right)\right|^2$, where $h_n\left(m\right)$ is the instantaneous gain of the channel between AP~$n$ and $\textnormal{UT}_0$. $H_n\left(r_n\right)$ can be expressed as a three-factor product:
\begin{equation}
H_n\left(r_n\right)=H_{\textnormal{pl},n}\left(r_n\right) G_{\textnormal{f},n} G_{\textnormal{s},n},
\label{Hnrn_Hpln_Gfn_Gsn}
\end{equation}
where $r_n$ represents the distance between $\textnormal{UT}_0$ and AP~$n$ (distances $r_n$ are functions of $\textnormal{UT}_0$'s position within its cell), and $H_{\textnormal{pl},n}\left(r_n\right)$, $G_{\textnormal{f},n}$ and $G_{\textnormal{s},n}$ represent the path loss, multipath Rayleigh fading and shadowing components respectively. We now further describe these last three components. 

The (deterministic) path loss $H_{\textnormal{pl},n}\left(r_n\right)$ diminishes as the distance $r_n$ between $\textnormal{UT}_0$ and AP~$n$ increases, based on the common power law \cite{Goldsmith2005}
\begin{equation}
H_{\textnormal{pl},n}\left(r_n\right) = K\left(\frac{d_0}{r_n}\right)^\gamma,
\label{Hpln_K_d0_rn_gamma}
\end{equation}
where $K=\left(c/\left(4\pi f d_0\right)\right)^2$ is a dimensionless constant, with $c$ being the speed of light, $f$, the operating frequency, and $d_0$, a reference distance for the antenna far-field; and $\gamma$ represents the path loss exponent. In order to make our study independent from the antenna characteristics and the cell size, we rewrite~\eqref{Hpln_K_d0_rn_gamma} under the following form:
\begin{equation}
H_{\textnormal{pl},n}\left(r_n\right) = K \left(\frac{d_0}{d_{\textnormal{ref}}}\right)^\gamma \left(\frac{d_{\textnormal{ref}}}{r_n}\right)^\gamma,
\label{Hplnrn_K_d0_dref_gamma_}
\end{equation}
where $d_{\textnormal{ref}}$ is a reference distance, and we introduce the \emph{normalized path loss} $G_{\textnormal{pl},n}\left(r_n\right)$, defined as follows:
\begin{equation}
G_{\textnormal{pl},n}\left(r_n\right) = \left(\frac{d_{\textnormal{ref}}}{r_n}\right)^\gamma. 
\label{Gpln_rn_}
\end{equation}
From~\eqref{Hplnrn_K_d0_dref_gamma_} and~\eqref{Gpln_rn_}, we establish the following relationship:
\begin{equation}
G_{\textnormal{pl},n}\left(r_n\right) = \frac{1}{K\left(\frac{d_0}{d_{\textnormal{ref}}}\right)^\gamma} H_{\textnormal{pl},n}\left(r_n\right).
\label{Gplnrn_1_K_}
\end{equation}
In a similar manner, we define the \emph{normalized instantaneous power gain} $G_n\left(r_n\right)$ as follows:
\begin{align}
G_n\left(r_n\right) & = \frac{1}{K\left(\frac{d_0}{d_{\textnormal{ref}}}\right)^\gamma} H_n\left(r_n\right) \nonumber \\
 & = G_{\textnormal{pl},n}\left(r_n\right) G_{\textnormal{f},n} G_{\textnormal{s},n}, 
\label{Gnrn_1_K_d0_dref_}
\end{align}
where~\eqref{Gnrn_1_K_d0_dref_} derives from~\eqref{Hnrn_Hpln_Gfn_Gsn} and~\eqref{Gplnrn_1_K_}. 

\subsection{\label{appendix1}Computation of moments for one interferer}
We find the closed form expression of the $k$th-order moment $\mathbb{E}\left\{\left(G_n\right)^k\right\}$ of the statistical distribution of the interference gain $G_n$ (one interfering cell). We have: 
\begin{align}
\mathbb{E}\left\{\left(G_n\right)^k\right\} &=\mathbb{E}\left\{\left(G_{\textnormal{f},n}G_{\textnormal{s},n}\right)^k\right\} \nonumber \\
&=\mathbb{E}\left\{\left(G_{\textnormal{f},n}\right)^k\right\} \mathbb{E}\left\{\left(G_{\textnormal{s},n}\right)^k\right\} \label{EGnk_EGfnk_EGsnk},
\end{align} 
where~\eqref{EGnk_EGfnk_EGsnk} follows from the independance property of the r.v.'s $G_{\textnormal{f},n}$ and $G_{\textnormal{s},n}$. As~$G_{\textnormal{f},n}$ is exponentially distributed with unit mean, its $k$th-order moment is given by:
\begin{equation}
\mathbb{E}\left\{\left(G_{\textnormal{f},n}\right)^k\right\}=k! \label{EGfnk_factk}
\end{equation}
As for~$G_{\textnormal{s},n}$, it has a lognormal distribution with parameters $-\sigma_{\textnormal{dB}}/2$ and $\sigma_{\textnormal{dB}}$; its raw moment can be written:
\begin{equation}
\mathbb{E}\left\{\left(G_{\textnormal{s},n}\right)^k\right\}=\exp\left(k(k-1) \frac{\sigma_{\textnormal{dB}}^2}{2}\right). \label{EGsnk_exp_}
\end{equation}•
Replacing~\eqref{EGfnk_factk} and~\eqref{EGsnk_exp_} in~\eqref{EGnk_EGfnk_EGsnk} leads to~\eqref{muk_k_exp_k_}. 

\subsection{\label{appendix2}Computation of moments for multiple interferers}
We establish the analytical formula of the $k$th-order moment $\mathbb{E}\left\{G^k\right\}$ of the statistical distribution of the interference gain $G$ (multiple interferers). Using approximation~\eqref{Gn_lambdan_Gfn_Gsn}, we can write:
\begin{align}
\mathbb{E}\left\{G^k\right\} &=\mathbb{E}\left\{\left(\sum_{n=1}^{N}\lambda_n G_{\textnormal{f},n} G_{\textnormal{s},n} \right)^k\right\} \nonumber \\
&=\mathbb{E}\left\{\sum_{\mathbf{a}:|\mathbf{a}|=k}\frac{k!}{\mathbf{a}!}Z^\mathbf{a} \right\}, \label{EGk_E_sum_aak_}
\end{align}
where the following notation is used:
\begin{itemize}
\item $\mathbf{a}=\left(\alpha_1,\alpha_2,\dots,\alpha_N\right)$, $\alpha_n\in \mathbb{N}$, $n=1,2,\dots,N$, is an $N$-dimensional vector whose sum of components is 
\begin{equation*}
|\mathbf{a}|=\sum_{n=1}^{N}\alpha_n ;
\end{equation*}
\item the multifactorial $\mathbf{a}!$ is such that 
\begin{equation*}
\mathbf{a}! = \prod_{n=1}^{N} \left(\alpha_n!\right) ;
\end{equation*}
\item the variable $Z^\mathbf{a}$ is defined as follows:
\begin{equation*}
Z^\mathbf{a} =\left(\lambda_1 G_{\textnormal{f},1}G_{\textnormal{s},1}\right)^{\alpha_1}  \left(\lambda_2 G_{\textnormal{f},2}G_{\textnormal{s},2}\right)^{\alpha_2} \cdots
\left(\lambda_N G_{\textnormal{f},N}G_{\textnormal{s},N}\right)^{\alpha_N}. 
\end{equation*}
\end{itemize}
Using \eqref{EGnk_EGfnk_EGsnk}, we can further develop~\eqref{EGk_E_sum_aak_}, which gives~\eqref{EZk_factk_sum_}. 

\subsection{\label{appendix3}Computation of correction factors}
We determine the correction factors used in the MCP method described in Section~\ref{N_interferers}. Recall that the technique consists, for non-compelled links, in randomly selecting intervals from a subset containing only the $\mathcal{J}$ highest-probability (i.e., smallest-amplitude) intervals. But, as high-amplitude intervals never appear in this random process, small amplitudes get overweighted in non-compelled links, which must be compensated in compelled links, where small (resp. large) amplitudes need to be underweighted (resp. overweighted), in such a way that the $1$st-order sampled moment converges to its exact value. Thus, in order to satisfy the mean constraint, an underweighting multiplicative factor, denoted $f^-$, is applied to amplitudes of the $\mathcal{J}$ first intervals of compelled links; similarly, an overweighting multiplicative factor $f^+$ is applied to amplitudes of the last $N-\mathcal{J}$ intervals. We now compute these two correction factors. 

Let us first see how each interfering link contributes to the $1$st-order moment of the intercell interference gain $G$. For each compelled link~$n$, $n=1,\dots,M$, we can write\footnote{Note that, for the sake of simplification, each $P$-element interval is reduced to its center of mass ---~denoted $g_j$.}:
\begin{align}
\mathbb{E}\left\{G_n\right\} &= \sum_{j=1}^{J} \delta_j g_j \nonumber \\
&= \underset{A}{\underbrace{\sum_{j=1}^{\mathcal{J}} \delta_j g_j}} + 
\underset{B}{\underbrace{\sum_{j=\mathcal{J}+1}^{J} \delta_j g_j}}  \nonumber \\
&=1 \nonumber,
\end{align}
where $G_n=G_{\textnormal{f},n} G_{\textnormal{s},n}$ (approximation~\eqref{Gn_Gfn_Gsn}, with $\lambda_n=1$), and, by construction of the typical set~$\mathcal{S}_n^\ell$, $A+B=1$, $\forall \sigma_{\textnormal{dB}}$. For each non-compelled link~$n$, $n=M+1,\dots,N$, $G_n$'s mean is
\begin{align}
\mathbb{E}\left\{G_n\right\} &= \sum_{j=1}^{\mathcal{J}} \delta_j' g_j \nonumber \\
&<1, \nonumber
\end{align}
where the probability set $\left\{\delta_j'\right\}$ is given by~\eqref{deltaj_alpha_deltaj_0}. So, if no correction factors are introduced, the contribution of all (compelled and non-compelled) links to the intercell interference gain $G$ gives the following mean:
\begin{align}
\mathbb{E}\left\{G\right\} &= \sum_{n=1}^{M}\lambda_n\underset{=1}{\underbrace{\mathbb{E}\left\{G_n\right\}}} + 
\sum_{n=M+1}^{N}\lambda_n\underset{<1}{\underbrace{\mathbb{E}\left\{G_n\right\}}} \nonumber \\
&<\sum_{n=1}^{N}\lambda_n, \nonumber
\end{align}
where 
\begin{equation}
\sum_{n=1}^{N}\lambda_n=A\sum_{n=1}^{N}\lambda_n + B\sum_{n=1}^{N}\lambda_n
\label{sum_lambda_sum_lambda_} 
\end{equation}
is the exact mean~\eqref{nu1}. 

Let us now introduce the correction factors $f^-$ and $f^+$ into compelled links, as described previously. $G$'s $1$st-order moment ---~denoted $\mathbb{E}_{\textnormal{cor}}\left\{G\right\}$~--- then becomes: 
\begin{align}
\mathbb{E}_{\textnormal{cor}}\left\{G\right\} &=\sum_{n=1}^{M}\lambda_n \left( \sum_{j=1}^{\mathcal{J}} \delta_j f^- g_j +  \sum_{j=\mathcal{J}+1}^{J} \delta_j f^+ g_j \right) + \sum_{n=M+1}^{N}\lambda_n \sum_{j=1}^{\mathcal{J}}\alpha \delta_j g_j \nonumber \\
& =\sum_{n=1}^{M}\lambda_n \left(Af^- + Bf^+ \right) + \sum_{n=M+1}^{N}\lambda_n \alpha A \nonumber \\
&=A\left(f^- \sum_{n=1}^{M}\lambda_n + \alpha \sum_{n=M+1}^{N}\lambda_n \right)+ B f^+\sum_{n=1}^{M}\lambda_n. 
\label{Ecor_}
\end{align}
In order for both exact and actual means to be equivalent (i.e., \eqref{sum_lambda_sum_lambda_}$\equiv$\eqref{Ecor_}), we need to solve the following system:
\begin{equation*}
\left\{
\begin{aligned}
f^- \sum_{n=1}^{M}\lambda_n + \alpha \sum_{n=M+1}^{N}\lambda_n = \sum_{n=1}^{N}\lambda_n \\
f^+\sum_{n=1}^{M}\lambda_n = \sum_{n=1}^{N}\lambda_n 
\end{aligned}
\right.
\end{equation*}
which leads to
\begin{equation}
f^- = 1-\left(\alpha-1\right) \frac{\underset{n=M+1}{\overset{N}{\sum}}\lambda_n}{\underset{n=1}{\overset{M}{\sum}}\lambda_n} 
\label{fminus}
\end{equation}
\begin{equation}
f^+ =1+\frac{\underset{n=M+1}{\overset{N}{\sum}}\lambda_n}{\underset{n=1}{\overset{M}{\sum}}\lambda_n}. 
\label{fplus}
\end{equation}
Note that we have $f^+>1$ and, as $\alpha\gtrsim 1$, $f^-\lesssim1$.

\end{appendix}

\bibliographystyle{IEEEtran}
\bibliography{biblio_WCN}



\begin{table}
\noindent \centering{}\caption{\label{tab_lambdan}Average path losses~$\lambda_{n}$, $n=1,2,\dots,N$, defined by~\eqref{Gn_lambdan_Gfn}, in decreasing order of importance. Each index $m$ of column~$\textnormal{AP}_{m}$ corresponds to index of average path loss~$\lambda_{n}$ ($n\neq m$, in general).}
\begin{tabular}{>{\centering}m{1cm}>{\centering}m{2cm}>{\centering}m{1cm}>{\centering}m{1cm}>{\centering}m{2cm}>{\centering}m{1cm}}
\multicolumn{3}{c}{$\textnormal{FR}1$ ($N=18$)} & \multicolumn{3}{c}{$\textnormal{FR}3$ ($N=6$)}\tabularnewline
$n$ & $\lambda_{n}$ & $\textnormal{AP}_{m}$ & $n$ & $\lambda_{n}$ & $\textnormal{AP}_{m}$\tabularnewline
$1$ & $6.467$ & $1$ & $1$ & $0.568$ & $8$\tabularnewline
$2$ & $3.588$ & $2$ & $2$ & $0.426$ & $18$\tabularnewline
$3$ & $1.708$ & $6$ & $3$ & $0.307$ & $10$\tabularnewline
$4$ & $1.069$ & $3$ & $4$ & $0.219$ & $16$\tabularnewline
$5$ & $0.767$ & $5$ & $5$ & $0.178$ & $12$\tabularnewline
$6$ & $0.663$ & $4$ & $6$ & $0.158$ & $14$\tabularnewline
$7$ & $0.568$ & $8$ &  &  & \tabularnewline
$8$ & $0.426$ & $18$ &  &  & \tabularnewline
$9$ & $0.316$ & $7$ &  &  & \tabularnewline
$10$ & $0.307$ & $10$ &  &  & \tabularnewline
$11$ & $0.260$ & $9$ &  &  & \tabularnewline
$12$ & $0.219$ & $16$ &  &  & \tabularnewline
$13$ & $0.188$ & $17$ &  &  & \tabularnewline
$14$ & $0.178$ & $12$ &  &  & \tabularnewline
$15$ & $0.158$ & $14$ &  &  & \tabularnewline
$16$ & $0.145$ & $11$ &  &  & \tabularnewline
$17$ & $0.118$ & $15$ &  &  & \tabularnewline
$18$ & $0.107$ & $13$ &  &  & \tabularnewline
\end{tabular}
\end{table}


\begin{table}
\noindent \centering{}\caption{\label{moments_single}Exact and approximated moments for one single interferer and for multiple interferers.}
\begin{tabular}{c>{\centering}p{2cm}>{\centering}p{2cm}>{\centering}p{2cm}>{\centering}p{2cm}}
 & \multicolumn{2}{>{\centering}p{5cm}}{no shadowing\\
 ($\sigma_{\textnormal{dB}}=0$ dB)} & \multicolumn{2}{>{\centering}p{5cm}}{intense shadowing\\
($\sigma_{\textnormal{dB}}=12$ dB)}\tabularnewline
 & exact & approximated & exact & approximated\tabularnewline
$\mathbb{E}\left\{\left(G_n\right)\right\}$ & $1$ & $1$ & $1$ & $0.990$\tabularnewline
$\mathbb{E}\left\{\left(G_n\right)^2\right\}$ & $2$ & $2$ & $4.138\cdot10^{3}$ & $1.119\cdot10^{3}$\tabularnewline
$\mathbb{E}\left\{\left(G_n\right)^3\right\}$ & $6$ & $6$ & $53.127\cdot10^{9}$ & $13.246\cdot10^{6}$\tabularnewline
$\mathbb{E}\left\{\left(G\right)\right\}$ (FR1) & $17.25$ & $17.10$ & $17.25$ & $17.08$\tabularnewline
$\mathbb{E}\left\{\left(G\right)\right\}$ (FR3) & $1.857$ & $1.857$ & $1.857$ & $1.855$\tabularnewline
\end{tabular}
\end{table}


\begin{table}
\noindent \centering{}\caption{\label{parameters_e_a_k_b}Coefficients $a_i$, $i=1,2,\dots,6$, of the empirical laws of parameters~$\eta$, $\alpha$, $k$, and~$\beta$ (FR1 and FR3 scenarios).}
\begin{tabular}{>{\centering}p{0.5cm}ccccccccccccc}
 & \multicolumn{6}{c}{$\textnormal{FR}1$} &  & \multicolumn{6}{c}{$\textnormal{FR}3$}\tabularnewline
 & $a_{1}$ & $a_{2}$ & $a_{3}$ & $a_{4}$ & $a_{5}$ & $a_{6}$ &  & $a_{1}$ & $a_{2}$ & $a_{3}$ & $a_{4}$ & $a_{5}$ & $a_{6}$\tabularnewline
$\eta$ & $4$ & $0$ & $1$ & $1$ & $1$ & $1$ &  & $0$ & $1$ & $1$ & $1$ & $1$ & $1$\tabularnewline
$\alpha$ & $0.93$ & $0.87$ & $65$ & $1$ & $7.2$ & $3.2$ &  & $0.38$ & $0.94$ & $39.90$ & $2.00$ & $8.30$ & $3.00$\tabularnewline
$k$ & $0.65$ & $2.18$ & $3.3$ & $0.39$ & $4.75$ & $2.06$ &  & $0$ & $12.70$ & $2.35$ & $2.07$ & $11.00$ & $6.47$\tabularnewline
$\beta$ & $0.04$ & $16.44$ & $13.45$ & $9$ & $6.35$ & $2.56$ &  & $1.81$ & $24.35$ & $3.60$ & $2.77$ & $1.77$ & $1.31$\tabularnewline
\end{tabular}
\end{table}

\begin{table}
\noindent \centering{}\caption{\label{parameter_xt}Coefficients $a_i$, $i=1,2,\dots,6$, of the empirical laws of parameter~$x_t$ (FR1 and FR3 scenarios).}
\begin{tabular}{cccccc}
 & $a_{1}$ & $a_{2}$ & $a_{3}$ & $a_{4}$ & $a_{5}$\tabularnewline
$\textnormal{FR}1$ & $61.56$ & $6.06$ & $1.84$ & $5.27$ & $2.51$\tabularnewline
$\textnormal{FR}3$ & $1.71$ & $5.10$ & $1.89$ & $6.40$ & $2.30$\tabularnewline
\end{tabular}
\end{table}

\clearpage


\end{document}